\documentclass[journal,twoside]{IEEEtran}

\usepackage{amsmath,amssymb,amsfonts}
\usepackage{mathrsfs}
\usepackage{textcomp}
\usepackage{microtype}

\makeatletter
\def\sciresult@one{1}
\newcommand{\sciresult}[2]{%
	\ensuremath{%
		\def\sciresult@mantissa{#1}%
		\ifx\sciresult@mantissa\sciresult@one
		\text{\rmfamily 10}^{\sciresult@exp#2\@nil}%
		\else
		\text{\rmfamily #1}\times \text{\rmfamily 10}^{\sciresult@exp#2\@nil}%
		\fi}}
\def\sciresult@exp#1#2\@nil{%
	\ifx#1-%
	-\text{\rmfamily #2}%
	\else
	\text{\rmfamily #1#2}%
	\fi}
\makeatother

\makeatletter
\def\section{\@startsection{section}{1}{\z@}
	{1.2ex plus 0.15ex minus 0.1ex}
	{0.45ex plus 0.1ex minus 0ex}
	{\normalfont\normalsize\centering\scshape}}
\def\subsection{\@startsection{subsection}{2}{\z@}
	{0.5ex plus 0.1ex minus 0.1ex}
	{0.35ex plus 0.1ex minus 0ex}
	{\normalfont\normalsize\itshape}}
\makeatother

\usepackage{graphicx}
\usepackage{booktabs}
\usepackage{array}
\usepackage{multirow}
\usepackage{makecell}
\usepackage{threeparttable}
\usepackage{stfloats}
\usepackage{colortbl}

\usepackage{algorithm}
\usepackage[noEnd=false,indLines=true]{algpseudocodex}
\tikzset{algpxIndentLine/.style={draw=black,line width=0.35pt}}

\usepackage{cite}
\usepackage{url}
\usepackage[colorlinks=true,linkcolor=blue,citecolor=blue,urlcolor=blue,hypertexnames=false]{hyperref}
\usepackage{soul}

\newcommand{\paperTitle}{Decentralized Continuous-Time Power Dispatch for Integrated Heat and Power Systems via Bernstein--Galerkin Equivalent Projection}

\newcommand{\nomenclatureIndexLabelWidth}{$l, \setL$}
\newcommand{\nomenclatureParamLabelWidth}{$\Delta t,n_t,n_s$}
\newcommand{\nomenclatureVarLabelWidth}
{$\boldsymbol\sigma_{\rm eq}^{(r)},
	\boldsymbol\sigma_{\rm ineq}^{(r)}$}

\newcommand{\R}{\mathbb{R}}
\newcommand{\setK}{\mathcal{K}}
\newcommand{\setL}{\mathcal{L}}

\newcommand{\standalonesubsubsection}[1]{%
	\par\refstepcounter{subsubsection}%
	\noindent{\normalfont\normalsize\itshape \thesubsubsectiondis\ #1}\par\nobreak
	\indent\ignorespaces
}
\renewcommand{\IEEEiedlistdecl}{%
	\setlength{\labelsep}{0.5em}
	\setlength{\itemsep}{0.1em}
	\setlength{\parsep}{0pt}
	\setlength{\topsep}{0pt}
}

\begin{document}
	\pagenumbering{arabic}
	\bstctlcite{IEEEexample:BSTcontrol}
	\title{\paperTitle}
	
	\author{Jie~Deng,
		Zhigang~Li,~\IEEEmembership{Senior Member, IEEE},
        J.~H.~Zheng,~\IEEEmembership{Member, IEEE},
        and~Yue~Chen,~\IEEEmembership{Senior Member, IEEE}
	}
	
	
	\maketitle
	\begin{abstract}
		Decentralized power dispatch for integrated heat and power systems (IHPSs) involves coordinating electric power systems (EPSs) and district heating networks (DHNs) while preserving privacy. The existing methods rely mainly on discrete-time DHN models and constant-flow operations, limiting the ability to exploit thermal flexibility. Enabling variable-flow and variable-temperature (VF-VT) operations requires thermal--hydraulic coupling to be addressed together with continuous-time thermal dynamics. However, the simultaneous variations in mass flows and temperature lead to a nonconvex IHPS model, for which conventional decomposition methods are difficult to apply. This paper proposes a decentralized continuous-time dispatch framework using Bernstein--Galerkin equivalent projection. For a prescribed mass flow trajectory, DHN thermal dynamics are formulated in the Bernstein space and projected onto boundary variables, yielding an equivalent feasible-region model without disclosing the internal DHN topology or states. The equivalent model and DHN subproblem are derived from a unified Bernstein--Galerkin formulation, ensuring a consistent thermal state representation throughout the EPS--DHN coordination procedure. A safeguarded Anderson prediction scheme further accelerates the alternating coordination process by using historical updates and fixed-point residuals to predict the next input without additional subproblem solving steps. Numerical results show the proposed method preserves thermal state consistency, exploits VF-VT flexibility, and improves economic and computational performance.
	\end{abstract}
	
	\vspace{-0.5em}
	
	\section*{Nomenclature}
	\addcontentsline{toc}{section}{Nomenclature}
	
	\subsection{Indices and Sets}
	\begin{IEEEdescription}[\IEEEusemathlabelsep \IEEEsetlabelwidth{\nomenclatureIndexLabelWidth}]
		\item[$j$] Index of prescribed mass flow trajectories.
		\item[$k, \setK$] Index and set of scheduling intervals, respectively.
		\item[$l, \setL$] Index and set of DHN pipelines, respectively.
		\item[$r$] Index of coordination iterations.
	\end{IEEEdescription}
	
	\subsection{Parameters, Matrices and Functions}
	\begin{IEEEdescription}[
		\IEEEusemathlabelsep
		\IEEEsetlabelwidth{\nomenclatureParamLabelWidth} ]
		\item[$\alpha_l, L_l$]
		Heat-loss coefficient and length of pipeline $l$, respectively.
		
		\item[$\boldsymbol A_{(\cdot)}^j$]
		Flow-evaluated thermal equation matrices of a pipeline.
		
		\item[$\boldsymbol B_n,\boldsymbol D_{n_t}^{\rm B}$]
		Bernstein basis vector and temporal derivative matrix, respectively.
		
		\item[$\boldsymbol C_{(\cdot)}^{\rm B},\boldsymbol b_{(\cdot)}^{\rm B}$]
		Coefficient matrices and right-hand-side vectors in the compact IHPS formulation, respectively.
		
		\item[$\Delta t,n_t,n_s$]
		Scheduling interval length and temporal and spatial Bernstein degrees, respectively.
		
		\item[$\boldsymbol F^{\rm ST}$]
		Spatiotemporal load matrix associated with the ambient temperature term.
		
		\item[$\boldsymbol{\it \Gamma}_{\rm CHP}^{\rm B}(\cdot)$]
		Operating function of CHP units.
		
		\item[$\boldsymbol H_{(\cdot)}^j$]
		Coefficient matrices in the pipeline outlet temperature equivalent.
		
		\item[$\boldsymbol K^\chi,\boldsymbol M^\chi$]
		Galerkin derivative and mass matrices, respectively.
		
		\item[$\Omega_{\rm H}^{{\rm B}, j}$]
		Projected feasible region of the DHN.
		
		\item[$\boldsymbol{\it \Pi}_{\rm pump}^{\rm B}(\cdot)$]
		Power consumption function of water pumps.
		
		\item[$\boldsymbol{\it \Psi}_{\Delta{\rm pr}}^{\rm B}(\cdot)$]
		Pressure loss function of DHN pipelines.
		
		\item[$T_{\rm amb}$]
		Ambient temperature.
		
		\item[$\widehat{\boldsymbol x}_{(\cdot)}^{{\rm B}, j}$]
		Constant vectors in the internal state equivalents.
		
		\item[$\boldsymbol Y_{(\cdot)}^j$]
		Internal state reconstruction matrices.
	\end{IEEEdescription}
	
	\subsection{Variables}
	\begin{IEEEdescription}[
		\IEEEusemathlabelsep
		\IEEEsetlabelwidth{\nomenclatureVarLabelWidth}
		]
		\item[$\boldsymbol\sigma_{\rm eq}^{(r)},
		\boldsymbol\sigma_{\rm ineq}^{(r)}$]
		Equality and inequality slack vectors in Stage I at iteration $r$, respectively.
		
		\item[$\boldsymbol\varpi^{(r)}$]
		Coordination vector at iteration $r$.
		
		\item[$\boldsymbol x_{\rm E}^{\rm B},
		\boldsymbol x_{\rm H}^{\rm B}$]
		Internal variable vectors of the EPS and DHN in the Bernstein space, respectively.
		
		\item[$\boldsymbol x_{\rm H,pr}^{\rm B},
		\boldsymbol x_{\rm H,T}^{\rm B}$]
		DHN pressure head and thermal state variable vectors in the Bernstein space, respectively.
		
		\item[$\boldsymbol x_{\rm m}^{\rm B}$]
		DHN mass flow variable vector in the Bernstein space.
		
		\item[$\boldsymbol x_{\tau,l,k}^{\rm B}$]
		Spatiotemporal temperature variable vector of pipeline $l$
        in interval $k$ in the Bernstein space.
		
		\item[$\boldsymbol z_{\rm E}^{\rm B},
		\boldsymbol z_{\rm H}^{\rm B}$]
		Boundary variable vectors associated with the EPS and DHN in the Bernstein space, respectively.
		
		\item[$\boldsymbol\zeta^{(r)},
		\boldsymbol\zeta_{\rm sc}^{(r)}$]
		Fixed-point residual and its scaled form at iteration $r$,
        respectively.
	\end{IEEEdescription}
	
	\section{Introduction}
	\IEEEPARstart{I}{ntegrated} heat and power systems (IHPSs) couple electric power systems (EPSs) with district heating networks (DHNs) to exploit the cross-sector flexibility derived from combined heat and power (CHP) units, heat pumps, and the inherent thermal inertia of the DHNs \cite{khatibi2021exploiting,xue2021reconfiguration}.
	In practice, the EPS and DHN of an IHPS are generally owned and operated by different entities with proprietary system data and distinct economic interests \cite{cao2019decentralized,zhu2026fully}. This motivates the decentralized coordination of EPS and DHN operations without requiring their private information to be fully disclosed.
	
	The existing decentralized approaches for solving IHPS power dispatch problems can be broadly classified into decomposition methods \cite{lu2020high,du2025globally}
and equivalent methods \cite{zheng2021dynamic}. The tractability of these methods depends on the IHPS model, which is affected by the presumed operating mode of the DHN. Under the constant-flow and variable-temperature (CF-VT) operation mode, the heat load demand is satisfied solely through temperature regulation with fixed mass flow rates. This operation mode simplifies the thermal--hydraulic coupling of the DHN and results in a linear DHN model, thereby facilitating the decomposition of the IHPS power dispatch problem into linear subproblems. However, this scheme shrinks the feasible DHN operation region and limits the full thermal flexibility that is attainable in coordinated power dispatch \cite{qiu2023decentralized,chen2021integrated}.
	
	Under the variable-flow and variable-temperature (VF-VT) operation mode, the heat load demand is supplied through the joint regulation of mass flow rates and temperatures \cite{qin2022increasing}, enhancing the operational flexibility of the DHN by releasing the fixed-flow restriction. However, the joint optimization of hydraulic and thermal regimes generally yields a nonconvex DHN model with nonlinear thermal--hydraulic coupling \cite{wu2022distributionally}. The existing decomposition-based formulations under the CF-VT operation mode strongly rely on convexity to obtain tractable subproblems and establish convergence guarantees \cite{huang2017coordinated,xue2020coordinated}. The convex structure underlying these formulations is generally lost under the VF-VT operation mode, necessitating a coordination paradigm that preserves feasibility without relying on model convexity.
	
	Equivalent-based coordination methods address this challenge by eliminating the internal states and projecting the feasible region onto the boundary variables. The resulting DHN equivalent model can be passed to the EPS operator without disclosing internal parameters. General equivalent-projection algorithms have been developed for the decentralized optimization of power systems \cite{tan2024parti,tan2024partii}. For decentralized IHPS power dispatch, equivalent dynamic models of DHNs have been established to represent their internal thermal states as functions of boundary variables \cite{zheng2021noniterative,zheng2022distributed}. The framework proposed in \cite{zheng2023distributed} extends equivalent-based coordination to the VF-VT operation mode by iteratively updating the DHN equivalent model to be embedded in the EPS subproblem.
	
	The existing DHN equivalent models for achieving decentralized IHPS power dispatch are primarily formulated in discrete time profiles, whereas the heat transport in DHNs is governed by partial differential equations (PDEs) and evolves continuously in time and space \cite{yang2020equivalent}. For computational convenience under the VF-VT operation mode, existing schemes use a discrete-time surrogate or quasidynamic model in the DHN subproblem together with a separately reconstructed DHN equivalent in the EPS subproblem \cite{li2015ComebinedHeat,zheng2023distributed}, yielding potentially inaccurate representations of the thermal dynamics. Continuous-time optimization represents temporal trajectories and spatiotemporal fields through basis function coefficients, and it has been applied to EPS scheduling \cite{parvania2016unit} and spatiotemporal energy flow modeling \cite{zheng2021energy,zhou2023function}. However, these studies focused on centralized optimization. A unified continuous-time framework for deriving an equivalent model of a DHN from its heat-transport PDEs while preserving thermal consistency remains to be developed for decentralized dispatch of IHPSs.
	
	The flow dependence of a DHN equivalent model introduces an outer coordination under the VF-VT operation mode. At each iteration, the equivalent model is reconstructed from the updated mass flow and embedded in the EPS power dispatch problem, whose solution provides the values of the boundary variables used in the DHN subproblem of the next iteration \cite{zheng2023distributed}. Since the equivalent model varies with the mass flow trajectory, this alternating procedure is essentially a nonlinear fixed-point iteration that is generally computationally intensive. An efficient alternating coordination scheme with residual-based fixed-point acceleration \cite{walker2011anderson,saad2025acceleration} is urgently needed for  continuous-time decentralized power dispatch in IHPSs.
	
	To address the above challenges, a decentralized continuous-time power dispatch method for IHPSs under the VF-VT operation mode is proposed in this paper using the Bernstein--Galerkin equivalent projection, with the following contributions.
	
	1) A continuous-time Bernstein--Galerkin DHN equivalent model is developed for decentralized IHPS dispatch under the VF-VT operation mode. For a prescribed mass flow trajectory, the heat-transport PDEs of the DHN are represented in the Bernstein space, and its internal thermal states are eliminated through equivalent projection. The resulting feasible-region model is expressed only by EPS--DHN boundary variables, enabling decentralized coordination without revealing the topology, parameters, or internal temperature profiles of the DHN while retaining continuous-time thermal dynamics.
	
	2) A unified continuous-time Bernstein--Galerkin formulation is established for both DHN equivalent modeling and DHN subproblem optimization. The equivalent model and the DHN subproblem are derived from the same Bernstein--Galerkin discretization of heat-transport dynamics. This avoids inconsistencies caused by separately constructed discrete-time or quasidynamic equivalents and DHN subproblems, maintains consistent thermal state trajectories, and preserves the feasibility of the DHN throughout the decentralized coordination process.
	
	3) A safeguarded Anderson prediction scheme is proposed to accelerate the alternating EPS--DHN coordination. Under the VF-VT operation mode, the flow-dependent DHN equivalent makes the coordination process a nonlinear fixed-point iteration. The proposed scheme uses historical alternating updates and fixed-point residuals to predict the next coordination input without additional EPS or DHN subproblem solving steps. A safeguard mechanism rejects unreliable extrapolations, improving robustness and computational efficiency.
	
	The remainder of this paper is organized as follows. Section~II formulates the continuous-time IHPS power dispatch problem and presents the EPS, DHN and boundary variables in the Bernstein space. Section~III develops the Bernstein--Galerkin equivalent projection of the DHN under prescribed mass flow trajectories. Section~IV presents the decentralized power dispatch algorithm with Anderson prediction. Section~V presents case studies. Section~VI concludes this paper.
	
	\section{Continuous-Time Formulation of Integrated Heat and Power Systems}
	Continuous-time IHPS power dispatch involves continuous temporal trajectories and spatiotemporal thermal fields, resulting in an infinite-dimensional functional problem. To obtain a tractable finite-dimensional formulation, these trajectories and fields are represented in the Bernstein space with their continuous-domain expressions preserved.
	
	Section~II-A introduces the Bernstein representations of temporal trajectories and spatiotemporal thermal fields. Section~II-B partitions the variables into EPS-side, DHN-side, and boundary components. Section~II-C presents the compact IHPS power dispatch formulation.
	
	\subsection{Representations of Continuous Variables in Bernstein Space}
	Before formulating the continuous-time IHPS power dispatch problem, the variables are defined in the Bernstein space. Temporal trajectories and spatiotemporal fields are parameterized by Bernstein coefficients while their continuous domains are retained.
	
	For a normalized coordinate $\eta\in[0,1]$ and a polynomial degree $n\in\mathbb{N}_0$, the Bernstein basis functions and the associated basis vector are defined as follows:
	\begin{align}
		B_{i,n}(\eta)&=\tbinom{n}{i}\eta^i(1-\eta)^{n-i},
		\; i=0,\ldots,n,\; 
		\label{eq:bernstein_basis}\\
		\boldsymbol{B}_{n}(\eta) &= [B_{0,n}(\eta),B_{1,n}(\eta),\ldots,B_{n,n}(\eta)]^\top.
		\label{eq:bernstein_vector}
	\end{align}
	
	Since the Bernstein basis is defined over $[0,1]$, the physical temporal and spatial coordinates are normalized as follows:
	\begin{equation}
		\begin{aligned}
			&\vartheta=(t-t_{k-1})/\Delta t,
			\; \vartheta\in[0,1],\; k\in\setK,\\
			&\quad \xi=s/L_l,\; \xi\in[0,1],\; l\in\setL,
		\end{aligned}
		\label{eq:normalization}
	\end{equation}
	where $\Delta t=t_k-t_{k-1}$ and $L_l$ denote the lengths of interval $k$ and pipeline $l$, respectively, and $\vartheta$ and $\xi$ are the corresponding normalized temporal and spatial coordinates, respectively.
	
	A generic time-dependent trajectory $g(\vartheta)$ over interval $k$ is described as follows:
	\begin{equation}
		\begin{aligned}
			& g(\vartheta) = (\boldsymbol{g}_{k}^{\rm B})^{\top}\boldsymbol{B}_{n_t}(\vartheta),
			\quad \forall n_t\in\mathbb{N}_0,\; k\in\setK, \\
			& \; \boldsymbol{g}_{k}^{\rm B} = [g_{k,0}^{\rm B},g_{k,1}^{\rm B},\ldots,g_{k,n_t}^{\rm B}]^\top
			\in\R^{n_t+1},
		\end{aligned}
		\label{eq:trajectory_rep}
	\end{equation}
	where $\boldsymbol{g}_{k}^{\rm B}$ is the Bernstein coefficient vector in interval $k$.
	
	The temporal differential and integral operators appearing in the continuous-time formulation can be evaluated directly from the Bernstein coefficients. With $t=t_{k-1}+\Delta t\,\vartheta$, these operators are expressed as follows:
	\begin{align}
		& 
        {\rm d}g(t)/{\rm d}t
		= 
		(\boldsymbol{D}_{n_t}^{\rm B}\boldsymbol{g}_{k}^{\rm B})^{\top}\boldsymbol{B}_{n_t-1}(\vartheta)/\Delta t,\;
        k\in\setK
		\label{eq:derivative}
		\\
		&\;\sum\nolimits_{k\in\setK}
		{\textstyle\int\nolimits_{t_{k-1}}^{t_k}}
        g(t)\,dt
		= \sum\nolimits_{k\in\setK}
		({\Delta t}/{n_t+1})
        \cdot\boldsymbol{1}_{n_t+1}^{\top}\boldsymbol{g}_{k}^{\rm B},
		\label{eq:integral}
	\end{align}
	where $\boldsymbol{D}_{n_t}^{\rm B}$ is the derivative matrix of the Bernstein basis. 
	
	For the thermal dynamics of the DHN, the temperature field of pipeline $l$ over interval $k$ is denoted by a spatiotemporal Bernstein basis as follows:
	\begin{equation}
		T_{l,k}(\xi,\vartheta)=\boldsymbol{B}_{n_s}^{\top}(\xi)\boldsymbol{T}_{l,k}^{\rm B}\boldsymbol{B}_{n_t}(\vartheta),
		\; l\in\setL,\; k\in\setK,
		\label{eq:temperature_surface}
	\end{equation}
	where $\boldsymbol{T}_{l,k}^{\rm B}\in\R^{(n_s+1)\times(n_t+1)}$ is the spatiotemporal coefficient matrix of the pipeline temperature. Hereafter, the superscript ${\rm B}$ indicates a Bernstein-space representation.
	
	\subsection{Partitioning of the Subsystem and Boundary Variables}
	The IHPS comprises an EPS and a DHN that are coupled through CHP units and water pumps. The supply and return networks in the DHN are radial, with the flow directions prescribed over the scheduling horizon. The decision variables are partitioned into internal and boundary components. The internal variables remain local to the corresponding subsystem, whereas the boundary variables characterize the coupling trajectories. The overall decision vector is defined as follows:
	\begin{equation}
		\boldsymbol{x}^{\mathrm{B}}
		=
		\operatorname{col}
		\{
		\boldsymbol{x}_{\rm E}^{\rm B},
		\boldsymbol{x}_{\rm H}^{\rm B},
		\boldsymbol{z}_{\rm E}^{\rm B},
		\boldsymbol{z}_{\rm H}^{\rm B}
		\},
		\label{eq:variable_partition}
	\end{equation}
	where $\operatorname{col}\{\cdot\}$ arranges its arguments into a column vector; $\boldsymbol{x}_{\rm E}^{\rm B}$ and $\boldsymbol{x}_{\rm H}^{\rm B}$ collect the internal variables of the EPS and DHN, respectively; and $\boldsymbol{z}_{\rm E}^{\rm B}$ and $\boldsymbol{z}_{\rm H}^{\rm B}$ collect their boundary variables.
	
	These four variable vectors are specified as follows:
	\begin{equation}
		\boldsymbol{x}_{\rm E}^{\rm B} = \operatorname{col}\{\boldsymbol{P}_{\rm G}^{\rm B},\boldsymbol{P}_{\rm W}^{\rm B},\boldsymbol{ru}^{\rm B},\boldsymbol{rd}^{\rm B}\},
		\label{eq:EPS_internal}
	\end{equation}
	\begin{equation}
		\boldsymbol{x}_{\rm H}^{\rm B} = \operatorname{col}\{\boldsymbol{x}_{\rm H,T}^{\rm B},\boldsymbol{x}_{\rm H,pr}^{\rm B},\boldsymbol{x}_{\rm m}^{\rm B}\},
		\label{eq:DHN_internal}
	\end{equation}
	\begin{equation}
		\boldsymbol{z}_{\rm E}^{\rm B} = \operatorname{col}\{\boldsymbol{P}_{\rm CHP}^{\rm B},\boldsymbol{P}_{\rm PUMP}^{\rm B}\},
		\label{eq:EPS_boundary}
	\end{equation}
	\begin{equation}
		\boldsymbol{z}_{\rm H}^{\rm B} = \operatorname{col}\{\boldsymbol{Q}_{\rm CHP}^{\rm B},\boldsymbol{T}_{\rm GS}^{\rm B},\boldsymbol{T}_{\rm GR}^{\rm B},\boldsymbol{pr}_{\rm ref}^{\rm B}\},
		\label{eq:DHN_boundary}
	\end{equation}
	where $\boldsymbol{P}_{\rm G}^{\rm B}$, $\boldsymbol{P}_{\rm W}^{\rm B}$, $\boldsymbol{ru}^{\rm B}$, and $\boldsymbol{rd}^{\rm B}$ denote the non-CHP power generation output, wind power output, upward reserve, and downward reserve, respectively. $\boldsymbol{x}_{\rm H,T}^{\rm B}$, $\boldsymbol{x}_{\rm H,pr}^{\rm B}$, and $\boldsymbol{x}_{\rm m}^{\rm B}$ collect the nodal temperatures, the pressure heads at the nonreference nodes and the mass flows, respectively.
	$\boldsymbol{P}_{\rm CHP}^{\rm B}$, $\boldsymbol{P}_{\rm PUMP}^{\rm B}$, $\boldsymbol{Q}_{\rm CHP}^{\rm B}$, $\boldsymbol{T}_{\rm GS}^{\rm B}$, $\boldsymbol{T}_{\rm GR}^{\rm B}$, and $\boldsymbol{pr}_{\rm ref}^{\rm B}$ denote the CHP power output, the pump power consumption, the CHP heat output, the heat source supply temperature, the heat source return temperature, and the pressure head of the reference node, respectively.
	
	\subsection{Compact Model of the IHPS Power Dispatch Problem}
	The Bernstein representations and coefficient relations described in \eqref{eq:trajectory_rep}--\eqref{eq:temperature_surface} are used to denote the objective and operating requirements of the continuous-time IHPS power dispatch process using Bernstein coefficients, yielding the following finite-dimensional nonlinear optimization problem:
	\begin{align}
		\mathrm{P}_0:
		&\min_{\boldsymbol{x}_{\rm E}^{\rm B},\boldsymbol{x}_{\rm H}^{\rm B},
			\boldsymbol{z}_{\rm E}^{\rm B},\boldsymbol{z}_{\rm H}^{\rm B}}\,
		J_{\rm op}^{\rm B}
		(\boldsymbol{x}_{\rm E}^{\rm B},
		\boldsymbol{z}_{\rm E}^{\rm B},
		\boldsymbol{z}_{\rm H}^{\rm B})
		\label{eq:P0_obj}
		\\
		\mathrm{s.t.}\;
		&
		\boldsymbol{C}_{\rm E,x}^{\rm B}
		\boldsymbol{x}_{\rm E}^{\rm B}
		+
		\boldsymbol{C}_{\rm E,z}^{\rm B}
		\boldsymbol{z}_{\rm E}^{\rm B}
		\leq
		\boldsymbol{b}_{\rm E}^{\rm B},
		\label{eq:P0_elec}
		\\
		&
		\boldsymbol{C}_{\rm th,x}^{\rm B}
		(\boldsymbol{x}_{\rm m}^{\rm B})
		\boldsymbol{x}_{\rm H,T}^{\rm B}
		+
		\boldsymbol{C}_{\rm th,z}^{\rm B}
		(\boldsymbol{x}_{\rm m}^{\rm B})
		\boldsymbol{z}_{\rm H}^{\rm B}
		=
		\boldsymbol{b}_{\rm th}^{\rm B},
		\label{eq:P0_heat_alg}
		\\
		&
		\mathcal{D}_{\rm H}^{\rm B}
		(\boldsymbol{x}_{\rm H,T}^{\rm B},
		\boldsymbol{x}_{\rm m}^{\rm B},
		\boldsymbol{z}_{\rm H}^{\rm B})
		=
		\boldsymbol{0},
		\label{eq:P0_heat_dyn}
		\\
		&
		\boldsymbol{C}_{\rm m}^{\rm B}
		\boldsymbol{x}_{\rm m}^{\rm B}
		=
		\boldsymbol{b}_{\rm m}^{\rm B},
		\quad
		\underline{\boldsymbol{x}}_{\rm m}^{\rm B}
		\leq
		\boldsymbol{x}_{\rm m}^{\rm B}
		\leq
		\overline{\boldsymbol{x}}_{\rm m}^{\rm B},
		\label{eq:P0_mass}
		\\
		&
		\boldsymbol{C}_{\rm pr,x}^{\rm B}
		\boldsymbol{x}_{\rm H,pr}^{\rm B}
		+
		\boldsymbol{C}_{\rm pr,z}^{\rm B}
		\boldsymbol{z}_{\rm H}^{\rm B}
		=
		\boldsymbol{\it \Psi}_{\Delta{\rm pr}}^{\rm B}
		(\boldsymbol{x}_{\rm m}^{\rm B}),
		\label{eq:P0_pressure}
		\\
		&
		\boldsymbol{\it \Gamma}_{\rm CHP}^{\rm B}
		(\boldsymbol{z}_{\rm E}^{\rm B},
		\boldsymbol{z}_{\rm H}^{\rm B})
		\leq
		\boldsymbol{0},
		\label{eq:P0_chp}
		\\
		&
		\boldsymbol{C}_{\rm pump,z}^{\rm B}
		\boldsymbol{z}_{\rm E}^{\rm B}
		=
		\boldsymbol{\it \Pi}_{\rm pump}^{\rm B}
		(\boldsymbol{x}_{\rm m}^{\rm B},
		\boldsymbol z_{\rm H}^{\rm B}),
		\label{eq:P0_pump}
	\end{align}
	where $\boldsymbol{C}_{(\cdot)}^{\rm B}$ and $\boldsymbol{b}_{(\cdot)}^{\rm B}$ denote the coefficient matrices and the associated constant vectors, respectively. The detailed formulation of $\mathrm{P}_0$ can be found in \cite{deng2026integrated}.
	
	The objective in \eqref{eq:P0_obj} minimizes the total operating cost of the IHPS, which is defined as $J_{\rm op}^{\rm B}
    =J_{\rm CU}^{\rm B}
    +J_{\rm CHP}^{\rm B}
    +J_{\rm W}^{\rm B}$, where $J_{\rm CU}^{\rm B}$, $J_{\rm CHP}^{\rm B}$, and $J_{\rm W}^{\rm B}$ denote the cost of non-CHP generation, CHP operation, and wind curtailment, respectively.
	
	Constraint \eqref{eq:P0_elec} represents the EPS operating constraints, including power balance, transmission limits, generation and wind power bounds, reserve requirements, and ramping rate limits.
	Equation \eqref{eq:P0_heat_alg} specifies the flow-dependent algebraic thermal constraints, including nodal temperature mixing, heat balances at the sources and loads, and pipeline node temperature continuity, whereas \eqref{eq:P0_heat_dyn} gives the Bernstein-space representation of the pipeline heat-transport constraints derived from the governing PDEs. Constraint \eqref{eq:P0_mass} enforces nodal mass balance and mass flow bounds, whereas \eqref{eq:P0_pressure} imposes the pressure-head drop equations, and $\boldsymbol{\it \Psi}_{\Delta{\rm pr}}^{\rm B}(\cdot)$ denotes the pressure loss function of DHN pipelines.
	Constraints \eqref{eq:P0_chp} and \eqref{eq:P0_pump} describe the CHP and pump coupling relations. $\boldsymbol{\it \Gamma}_{\rm CHP}^{\rm B}(\cdot)$ defines the feasible CHP operation regions, while \eqref{eq:P0_pump} equates the pump power in $\boldsymbol{z}_{\rm E}^{\rm B}$ to the hydraulic power requirement $\boldsymbol{\it \Pi}_{\rm pump}^{\rm B}(\cdot)$.
	
	Although the centralized formulation $\mathrm{P}_0$ is finite-dimensional, its solution requires the EPS and DHN operators to disclose proprietary operating data. To enable decentralized coordination without such disclosures, the internal thermal and hydraulic states of the DHN are represented by the boundary variables to be exchanged under a prescribed mass flow trajectory.
	
	\section{Bernstein--Galerkin Equivalent Projection for District Heating Networks} 
	In this section, a Bernstein--Galerkin equivalent projection is derived from the DHN constraints in $\mathrm{P}_0$. The projected feasible region of DHN operations is incorporated into the EPS subproblem and is constantly updated during the iterative decentralized coordination procedure.
	
	Section~III-A reformulates the thermal dynamics of the DHN in a Bernstein--Galerkin form. Section~III-B derives an equivalent representation of the thermal dynamics under the prescribed mass flow trajectory. Section~III-C constructs the thermal and hydraulic state equivalents and projects the feasible region of the DHN onto the boundary variables.
	
	\subsection{Bernstein--Galerkin Reformulation of the DHN Dynamics}
	Constructing the DHN equivalent requires a finite-dimensional expression of the continuous spatiotemporal heat transport. The Bernstein-- Galerkin method transforms the PDEs that govern the pipelines' thermal dynamics into algebraic equations. For pipeline $l$ over interval $k$, the temperature field satisfies the following heat transfer equation:
	\begin{equation}
		\frac{\partial T_l(s,t)}{\partial t}
		+v_l(t)\frac{\partial T_l(s,t)}{\partial s}
		+\alpha_l \left(T_l(s,t)-T_{\rm amb}\right)=0,
		\label{eq:pipe_pde}
	\end{equation}
	where $T_l(s,t)$ denotes the pipeline temperature and $v_l(t)$ is the flow velocity induced by $m_l(t)$. After applying the coordinate normalization process described in \eqref{eq:normalization}, the temporal and spatial derivatives are scaled by $1/\Delta t$ and $1/L_l$, respectively.
	
	The Galerkin condition requires the residual of the normalized PDE to be orthogonal to the Bernstein testing functions:
	\begin{equation}
		\begin{aligned}
			&{\textstyle\int\nolimits_{0}^{1}}
			{\textstyle\int\nolimits_{0}^{1}}
			\mathcal R_{l,k}(\xi,\vartheta)
			B_{p,n_s}(\xi)B_{q,n_t}(\vartheta)
			\,\mathrm d\xi\,\mathrm d\vartheta=0,
		\end{aligned}
		\label{eq:galerkin_condition}
	\end{equation}
	where $\mathcal R_{l,k}(\xi,\vartheta)$ denotes the residual of the normalized form of \eqref{eq:pipe_pde} over interval $k$, $q=1,\ldots,n_t$, while $p=1,\ldots,n_s$ for a supply pipeline and $p=0,\ldots,n_s-1$ for a return pipeline.
	
	Substituting the Bernstein representations of $T_{l,k}(\xi,\vartheta)$ and $m_{l,k}(\vartheta)$ into \eqref{eq:galerkin_condition} yields the Bernstein--Galerkin equations:
	\begin{equation}
		\begin{aligned}
			& \big[
			1/{\Delta t} \cdot
			\boldsymbol M^{\rm SP}
			\boldsymbol T_{l,k}^{\rm B}
			(\boldsymbol K^{\rm TE})^{\top}
			+
			v_l \boldsymbol K^{\rm SP} \boldsymbol T_{l,k}^{\rm B}
			\boldsymbol M^{\rm TE} 
			\\
			& \; 
			+ \alpha_l \boldsymbol M^{\rm SP} \boldsymbol T_{l,k}^{\rm B}
			\boldsymbol M^{\rm TE}
			- \alpha_l T_{\rm amb} \boldsymbol F^{\rm ST}
			\big]_{p,q}
			= 0,
		\end{aligned}
		\label{eq:gb_pipe_thermal}
	\end{equation}
	where $\boldsymbol T_{l,k}^{\rm B}\in\R^{(n_s+1)\times(n_t+1)}$ contains the pipeline temperature, and the indices $p$ and $q$ follow the testing ranges described in \eqref{eq:galerkin_condition}.
	The superscripts ${\rm SP}$, ${\rm TE}$, and ${\rm ST}$ indicate spatial, temporal, and spatiotemporal representations, respectively.
	
	For $\chi\in\{{\rm SP},{\rm TE}\}$, the constant Galerkin matrices and vectors in \eqref{eq:gb_pipe_thermal} are defined as follows:
	\begin{equation}
		\begin{aligned}
			[\boldsymbol M^\chi]_{p,q}
			&={\textstyle\int\nolimits_0^1}
			B_{p,n_\chi}(\eta_\chi)B_{q,n_\chi}(\eta_\chi)\,\mathrm d\eta_\chi,
			\\
			[\boldsymbol K^\chi]_{p,q}
			&={\textstyle\int\nolimits_0^1}
			B_{p,n_\chi}(\eta_\chi)
			[\mathrm d B_{q,n_\chi}(\eta_\chi)/\mathrm d\eta_\chi]\,\mathrm d\eta_\chi,
			\\
			[\boldsymbol f^\chi]_{p}
			&={\textstyle\int\nolimits_0^1}
			B_{p,n_\chi}(\eta_\chi)\,\mathrm d\eta_\chi,\quad
			\boldsymbol F^{\rm ST}
			=\boldsymbol f^{\rm SP}(\boldsymbol f^{\rm TE})^\top,
		\end{aligned}
		\label{eq:constant_galerkin_matrices}
	\end{equation}
	where $(\eta_{\rm SP},n_{\rm SP})=(\xi,n_s)$, $(\eta_{\rm TE},n_{\rm TE})=(\vartheta,n_t)$, and $p,q=0,\ldots,n_\chi$. $\boldsymbol F^{\rm ST}$ represents the spatiotemporal load matrix that is associated with the ambient temperature term.
	
	\subsection{Equivalent Representation of DHN Thermal Dynamics}
	Following the equivalent projection principle proposed in \cite{zheng2021dynamic,zheng2023distributed}, a relation for the outlet temperatures of pipelines is derived from the Bernstein--Galerkin equations in \eqref{eq:gb_pipe_thermal}. For a prescribed mass flow trajectory $\boldsymbol{x}_{\rm m}^{{\rm B},j}$, the corresponding flow velocities are fixed, rendering the Bernstein--Galerkin equations linear with respect to the temperature. Hereafter, the superscript $j$ indicates the evaluation performed at $\boldsymbol{x}_{\rm m}^{{\rm B},j}$.
	
	For pipeline $l$ in interval $k$, the temperature vectors are defined as $\boldsymbol{x}_{\tau,l,k}^{\rm B} =\operatorname{vec}(\boldsymbol{T}_{l,k}^{\rm B})$. 
	Vectorizing \eqref{eq:gb_pipe_thermal} and incorporating the inlet and initial conditions yields the following:
	\begin{equation}
		\begin{aligned}
			&\boldsymbol A_{{\rm ag},l,k}^{j}
			\boldsymbol x_{\tau,l,k}^{\rm B}
			= \operatorname{col}
			\big\{
			\boldsymbol f_{\tau,l,k}^{\rm B}, \,
			[\boldsymbol 0 \; \boldsymbol I_{n_t}] 
			\boldsymbol x_{\tau,{\rm in},l,k}^{\rm B}, \,
			\boldsymbol x_{\tau,0,l,k}^{\rm B}
			\big\},
			\\
			&\quad \boldsymbol A_{{\rm ag},l,k}^{j}
			= \operatorname{col}
			\big\{
			\boldsymbol A_{\tau,l,k}^{j},
			\boldsymbol E_{{\rm in},l}^{\rm B},
			\boldsymbol E_{0}^{\rm B}
			\big\},
		\end{aligned}
		\label{eq:pipe_augmented_relation}
	\end{equation}
	where $\boldsymbol E_{{\rm in},l}^{\rm B} =[\boldsymbol 0 \; \boldsymbol I_{n_t}]\otimes \boldsymbol e_{{\rm in},l}^{\top}$ and $\boldsymbol E_{0}^{\rm B} =\boldsymbol e_{0,t}^{\top}\otimes \boldsymbol I_{n_s+1}$ represent the inlet temperature and initial temperature, respectively. Here, $\boldsymbol e_{0,t}=[1,0,\ldots,0]^\top$, while $\boldsymbol e_{{\rm in},l}=[1,0,\ldots,0]^\top$ for a supply pipeline and $\boldsymbol e_{{\rm in},l}=[0,\ldots,0,1]^\top$ for a return pipeline.
	
	The corresponding Galerkin matrix and the ambient temperature source vector are given as follows:
	\begin{equation}
		\begin{aligned}
			& \boldsymbol A_{\tau,l,k}^j
			=\big[\frac{1}{\Delta t}
			\bigl(\boldsymbol K^{\rm TE}\otimes\boldsymbol M^{\rm SP}\bigr)
			+v_l^j
			\bigl[\bigl(\boldsymbol M^{{\rm TE},j}\bigr)^\top \\
			&\qquad \qquad \otimes\boldsymbol K^{\rm SP}\bigr]
			+\alpha_l
			\bigl[\bigl(\boldsymbol M^{\rm TE}\bigr)^\top
			\otimes\boldsymbol M^{\rm SP}\bigr]\big]_{p,q},
			\\
			& \quad \boldsymbol f_{\tau,l,k}^{\rm B}
			=\alpha_lT_{\rm amb}[\operatorname{vec}(\boldsymbol F^{\rm ST})]_{p,q}.
		\end{aligned}
		\label{eq:pipe_vectorized_terms}
	\end{equation}
	where the subscript $(p,q)$ retains the rows that are associated with the testing ranges in \eqref{eq:galerkin_condition}.
	
	The Bernstein--Galerkin equations and the associated inlet and initial conditions form a square system with a nonsingular coefficient matrix $\boldsymbol A_{{\rm ag},l,k}^{j}$. Solving \eqref{eq:pipe_augmented_relation} then yields the following affine representation of the pipeline temperature:
    \begin{align}
	&[
	\boldsymbol Y_{{\rm in},l,k}^{j}\;
	\boldsymbol Y_{0,l,k}^{j}\;
	\widehat{\boldsymbol x}_{\tau,l,k}^{{\rm B},j}
	]
	=
	(\boldsymbol A_{{\rm ag},l,k}^{j})^{-1}
	[
	\operatorname{col}
	\{\boldsymbol 0,[\boldsymbol 0\;\boldsymbol I_{n_t}],\boldsymbol 0\},
	\notag
	\\
	&\hspace{3em}
	\operatorname{col}
	\{\boldsymbol 0,\boldsymbol 0,\boldsymbol I_{n_s+1}\},
	\operatorname{col}
	\{\boldsymbol f_{\tau,l,k}^{\rm B},\boldsymbol 0,\boldsymbol 0\}
	],
	\label{eq:pipe_temperature_mapping}
	\\
	&\boldsymbol x_{\tau,l,k}^{\rm B}
	=
	\boldsymbol Y_{{\rm in},l,k}^{j}
	\boldsymbol x_{\tau,{\rm in},l,k}^{\rm B}
	+
	\boldsymbol Y_{0,l,k}^{j}
	\boldsymbol x_{\tau,0,l,k}^{\rm B}
	+
	\widehat{\boldsymbol x}_{\tau,l,k}^{{\rm B},j}.
	\label{eq:pipe_temperature_equivalent}
    \end{align}

	The outlet temperatures required by the algebraic constraints of the network are extracted from \eqref{eq:pipe_temperature_equivalent} as follows:
	\begin{equation}
		\begin{aligned}
			&\boldsymbol x_{\tau,{\rm out},l,k}^{\rm B} \! = \! \boldsymbol H_{{\rm in},l,k}^j \boldsymbol x_{\tau,{\rm in},l,k}^{\rm B} \!+\! \boldsymbol H_{0,l,k}^j \boldsymbol x_{\tau,0,l,k}^{\rm B} \!+\! \widehat{\boldsymbol x}_{\tau,{\rm out},l,k}^{{\rm B},j}, 
			\\
			& \quad \boldsymbol H_{{\rm in},l,k}^j = \boldsymbol R_{{\rm out},l} \boldsymbol Y_{{\rm in},l,k}^j, \;
			\boldsymbol H_{0,l,k}^j = \boldsymbol R_{{\rm out},l} \boldsymbol Y_{0,l,k}^j, \\
			& \quad \widehat{\boldsymbol x}_{\tau,{\rm out},l,k}^{{\rm B},j} = \boldsymbol R_{{\rm out},l} \widehat{\boldsymbol x}_{\tau,l,k}^{{\rm B},j},
		\end{aligned}
		\label{eq:pipe_outlet_equivalent}
	\end{equation}
	where $\boldsymbol R_{{\rm out},l} =\boldsymbol I_{n_t+1}\otimes \boldsymbol e_{{\rm out},l}^{\top}$ determines the outlet temperature. For a supply pipeline, $\boldsymbol e_{{\rm out},l}=[0,\ldots,0,1]^{\top}$; for a return pipeline, $\boldsymbol e_{{\rm out},l}=[1,0,\ldots,0]^{\top}$.
	
	Collecting the outlet temperature relations in \eqref{eq:pipe_outlet_equivalent} for all the pipelines $l\in\setL$ and intervals $k\in\mathcal K$ yields the following DHN-wide relation for the pipeline outlet temperatures:
	\begin{equation}
		\boldsymbol x_{\tau,{\rm out}}^{\rm B}
		=
		\boldsymbol H_{\rm in}^j
		\boldsymbol x_{\tau,{\rm in}}^{\rm B}
		+
		\boldsymbol H_{0}^j
		\boldsymbol x_{\tau,0}^{\rm B}
		+
		\widehat{\boldsymbol x}_{\tau,{\rm out}}^{{\rm B},j},
		\label{eq:global_pipe_outlet_equivalent}
	\end{equation}
	where $\boldsymbol H_{\rm in}^j$, $\boldsymbol H_{0}^j$, and $\widehat{\boldsymbol x}_{\tau,{\rm out}}^{{\rm B},j}$ are assembled from the corresponding $\boldsymbol H_{{\rm in},l,k}^{j}$, $\boldsymbol H_{0,l,k}^{j}$, and $\widehat{\boldsymbol x}_{\tau,{\rm out},l,k}^{{\rm B},j}$, respectively. 
	
	\subsection{Equivalent Projection of the Feasible Region of the DHN}
	
	\subsubsection{Thermal State Equivalent and Projection}
	To derive the thermal state equivalent, the DHN-wide pipeline outlet temperature relation given in \eqref{eq:global_pipe_outlet_equivalent} is incorporated into the algebraic thermal constraints described in \eqref{eq:P0_heat_alg}. The internal thermal variables are reordered and partitioned as $\boldsymbol x_{\rm H,T}^{\rm B} =\operatorname{col}\{ \boldsymbol x_{\tau,{\rm out}}^{\rm B}, \boldsymbol x_{{\rm H,T},{\rm rm}}^{\rm B}\}$, where $\boldsymbol x_{{\rm H,T},{\rm rm}}^{\rm B}$ collects all internal thermal variables other than the pipeline outlet temperature.
	
	At $\boldsymbol x_{\rm m}^{{\rm B},j}$, the state reconstruction equations in \eqref{eq:P0_heat_alg} are partitioned according to the above decomposition as follows:
	\begin{equation}
		\boldsymbol C_{\rm th,out}^{{\rm B},j}
		\boldsymbol x_{\tau,{\rm out}}^{\rm B}
		+
		\boldsymbol C_{\rm th,rm}^{{\rm B},j}
		\boldsymbol x_{{\rm H,T},{\rm rm}}^{\rm B}
		+
		\boldsymbol C_{\rm th,z}^{{\rm B},j}
		\boldsymbol z_{\rm H}^{\rm B}
		=
		\boldsymbol b_{\rm th}^{\rm B},
		\label{eq:alg_partition}
	\end{equation}
	where the coefficient matrices are associated with $\boldsymbol x_{\tau,{\rm out}}^{\rm B}$, $\boldsymbol x_{{\rm H,T},{\rm rm}}^{\rm B}$, and $\boldsymbol z_{\rm H}^{\rm B}$, respectively, and $\boldsymbol b_{\rm th}^{\rm B}$ is the associated right-hand vector.
	
	The inlet temperature of each pipeline is determined by the temperature at its inlet node. Collecting these relations across all the pipelines and intervals yields the following:
	\begin{equation}
		\boldsymbol x_{\tau,{\rm in}}^{\rm B}
		=
		\boldsymbol S_{{\rm in},{\rm rm}}
		\boldsymbol x_{{\rm H,T},{\rm rm}}^{\rm B}
		+
		\boldsymbol S_{{\rm in},{\rm z}}
		\boldsymbol z_{\rm H}^{\rm B},
		\label{eq:pipe_inlet_relation}
	\end{equation}
	where $\boldsymbol S_{{\rm in},{\rm rm}}$ and $\boldsymbol S_{{\rm in},{\rm z}}$ are selection matrices determined by the DHN topology. For pipeline $l$ in interval $k$, the corresponding submatrix satisfies $\left[ \boldsymbol S_{{\rm in},{\rm rm}} \quad \boldsymbol S_{{\rm in},{\rm z}} \right]_{l,k} = \left[\boldsymbol 0 \ \cdots\ \boldsymbol I_{n_t+1} \ \cdots\ \boldsymbol 0 \right]$, where $\boldsymbol I_{n_t+1}$ denotes the identity matrix, whose position in the submatrix corresponds to the inlet node temperature.
	
	Substituting \eqref{eq:pipe_inlet_relation} and \eqref{eq:global_pipe_outlet_equivalent} into \eqref{eq:alg_partition} yields the following relation between the remaining internal thermal variables and the DHN boundary variables:
	\begin{equation}
		\widetilde{\boldsymbol C}_{\rm th,rm}^{j}
		\boldsymbol x_{{\rm H,T},{\rm rm}}^{\rm B}
		+
		\widetilde{\boldsymbol C}_{\rm th,z}^{j}
		\boldsymbol z_{\rm H}^{\rm B}
		=
		\widetilde{\boldsymbol b}_{\rm th}^{j},
		\label{eq:reduced_heat_alg}
	\end{equation}
	where
	$\widetilde{\boldsymbol C}_{\rm th,rm}^{j}
	= \boldsymbol C_{\rm th,rm}^{{\rm B},j}
	+ \boldsymbol C_{\rm th,out}^{{\rm B},j}
	\boldsymbol H_{\rm in}^{j}
	\boldsymbol S_{{\rm in},{\rm rm}}$
	and
	$\widetilde{\boldsymbol C}_{\rm th,z}^{j}
	= \boldsymbol C_{\rm th,z}^{{\rm B},j}
	+ \boldsymbol C_{\rm th,out}^{{\rm B},j}
	\boldsymbol H_{\rm in}^{j}
	\boldsymbol S_{{\rm in},{\rm z}}$.
	The right-hand vector is
	$\widetilde{\boldsymbol b}_{\rm th}^{j}
	= \boldsymbol b_{\rm th}^{\rm B}
	- \boldsymbol C_{\rm th,out}^{{\rm B},j}
	(\boldsymbol H_{0}^{j}
	\boldsymbol x_{\tau,0}^{\rm B}
	+ \widehat{\boldsymbol x}_{\tau,{\rm out}}^{{\rm B},j})$.
	
	The radial DHN topology assumed in this paper satisfies the existence condition for the dynamic equivalent model of the DHN in \cite{zheng2021noniterative}, rendering $\widetilde{\boldsymbol C}_{\rm th,rm}^{j}$ nonsingular. The remaining internal thermal variables are then expressed as follows:
	\begin{equation}
		\boldsymbol x_{{\rm H,T},{\rm rm}}^{\rm B}
		=
		\boldsymbol Y_{\rm T}^{j}
		\boldsymbol z_{\rm H}^{\rm B}
		+
		\widehat{\boldsymbol x}_{\rm T}^{{\rm B},j},
		\label{eq:remaining_thermal_equivalent}
	\end{equation}
	where $\boldsymbol Y_{\rm T}^{j}= -(\widetilde{\boldsymbol C}_{\rm th,rm}^{j})^{-1} \widetilde{\boldsymbol C}_{\rm th,z}^{j}$ and $\widehat{\boldsymbol x}_{\rm T}^{{\rm B},j} = (\widetilde{\boldsymbol C}_{\rm th,rm}^{j} )^{-1}\widetilde{\boldsymbol b}_{\rm th}^{j}$.
	
	The thermal state equivalent for $\boldsymbol x_{\rm H,T}^{\rm B}$ in terms of the DHN boundary variables is then obtained from \eqref{eq:remaining_thermal_equivalent} and the pipeline outlet temperature relation given in \eqref{eq:global_pipe_outlet_equivalent}:
	\begin{equation}
		\boldsymbol x_{\rm H,T}^{\rm B}
		=
		\begin{bmatrix}
			\boldsymbol Y_{\tau,{\rm out}}^{j}\\
			\boldsymbol Y_{\rm T}^{j}
		\end{bmatrix}
		\boldsymbol z_{\rm H}^{\rm B}
		+
		\begin{bmatrix}
			\widehat{\boldsymbol x}_{\tau,{\rm z}}^{{\rm B},j}\\
			\widehat{\boldsymbol x}_{\rm T}^{{\rm B},j}
		\end{bmatrix}
		=
		\boldsymbol Y_{\rm H}^{j}
		\boldsymbol z_{\rm H}^{\rm B}
		+
		\widehat{\boldsymbol x}_{\rm H,T}^{{\rm B},j},
		\label{eq:thermal_all_equivalent}
	\end{equation}
	where
	$\boldsymbol Y_{\tau,{\rm out}}^{j}
	= \boldsymbol H_{\rm in}^{j}
	(\boldsymbol S_{{\rm in},{\rm rm}}
	\boldsymbol Y_{\rm T}^{j}
	+\boldsymbol S_{{\rm in},{\rm z}})$.
	The corresponding constant vector is 
	$\widehat{\boldsymbol x}_{\tau,{\rm z}}^{{\rm B},j}
	=\boldsymbol H_{\rm in}^{j}
	\boldsymbol S_{{\rm in},{\rm rm}}
	\widehat{\boldsymbol x}_{\rm T}^{{\rm B},j}
	+\boldsymbol H_{0}^{j}
	\boldsymbol x_{\tau,0}^{\rm B}
	+\widehat{\boldsymbol x}_{\tau,{\rm out}}^{{\rm B},j}$.
	
	The internal thermal variables satisfy the componentwise bounds $\underline{\boldsymbol x}_{\rm H,T}^{\rm B}\leq \boldsymbol x_{\rm H,T}^{\rm B} \leq \overline{\boldsymbol x}_{\rm H,T}^{\rm B}$. Substituting the thermal state equivalent in \eqref{eq:thermal_all_equivalent} projects these limits onto the DHN boundary variables:
	\begin{equation}
		\begin{bmatrix}
			\boldsymbol Y_{\rm H}^{j}\\
			-\boldsymbol Y_{\rm H}^{j}
		\end{bmatrix}
		\boldsymbol z_{\rm H}^{\rm B}
		\leq
		\begin{bmatrix}
			\overline{\boldsymbol x}_{\rm H,T}^{\rm B}
			-
			\widehat{\boldsymbol x}_{\rm H,T}^{{\rm B},j}\\
			-\underline{\boldsymbol x}_{\rm H,T}^{\rm B}
			+
			\widehat{\boldsymbol x}_{\rm H,T}^{{\rm B},j}
		\end{bmatrix}.
		\label{eq:projected_temperature_limits}
	\end{equation}
	
	Constraints~\eqref{eq:thermal_all_equivalent} and \eqref{eq:projected_temperature_limits} constitute the thermal constraints of the projected feasible region of the DHN.
	
	\subsubsection{Hydraulic State Equivalent and Projection}
	For the prescribed mass flow trajectory $\boldsymbol x_{\rm m}^{{\rm B},j}$, the pressure loss terms of the pipeline in \eqref{eq:P0_pressure} are fixed. For the supply and return networks, which are indexed by $\nu\in\{{\rm S},{\rm R}\}$, $\mathcal P_{\nu,h}$ denotes the oriented path from the reference node to the nonreference node $h$. The cumulative pressure losses along these paths are given by the following equation:
	\begin{equation}
		\begin{aligned}
			&\boldsymbol \ell_{\rm pr,\nu}^{{\rm B},j}
			\! = \! (\boldsymbol N_{\rm pr,\nu}\otimes\boldsymbol I_{n_t+1})
			\boldsymbol {\it \Psi}_{\Delta{\rm pr},\nu}^{\rm B}
			( \boldsymbol x_{\rm m}^{{\rm B},j} ),
			\\
			& \quad
			[\boldsymbol N_{\rm pr,\nu}]_{h,l}
			\! = \! \mathbb I (l\in\mathcal P_{\nu,h}),
		\end{aligned}
		\label{eq:path_pressure_loss}
	\end{equation}
	where $\boldsymbol\ell_{\rm pr,\nu}^{{\rm B},j}$ is the resulting cumulative pressure loss vector, $\boldsymbol N_{\rm pr,\nu}$ is the path incidence matrix, and $\mathbb I(\cdot)$ denotes the indicator function.
	
	Utilizing the reference node pressure head in $\boldsymbol z_{\rm H}^{\rm B}$, the pressure heads at the nonreference nodes are reconstructed as follows:
	\begin{equation}
		\boldsymbol{pr}_{\rm S, nr}^{\rm B}
		=
		\boldsymbol S_{\rm pr,S}^{\rm ref}
		\boldsymbol z_{\rm H}^{\rm B}
		-
		\boldsymbol \ell_{\rm pr,S}^{{\rm B},j},
		\;
		\boldsymbol{pr}_{\rm R, nr}^{\rm B}
		=
		\boldsymbol S_{\rm pr,R}^{\rm ref}
		\boldsymbol z_{\rm H}^{\rm B}
		+
		\boldsymbol \ell_{\rm pr,R}^{{\rm B},j},
		\label{eq:nonreference_pressure_reconstruction}
	\end{equation}
	where $\boldsymbol{pr}_{\nu,{\rm nr}}^{\rm B}$ represents the pressure heads located at the nonreference nodes of network $\nu$, and $\boldsymbol S_{\rm pr,\nu}^{\rm ref}$ is a selection matrix that is determined by the DHN topology. For nonreference node $h$ in interval $k$, the associated submatrix satisfies $[ \boldsymbol S_{\rm pr,\nu}^{\rm ref}]_{h,k} = [ \boldsymbol 0 \ \cdots\ \boldsymbol I_{n_t+1} \ \cdots\ \boldsymbol 0]$, where the position of $\boldsymbol I_{n_t+1}$ corresponds to the reference-node pressure head of network $\nu$.
	
	With $\boldsymbol x_{\rm H,pr}^{\rm B} =\operatorname{col}\{ \boldsymbol{pr}_{\rm S, nr}^{\rm B}, \, \boldsymbol{pr}_{\rm R, nr}^{\rm B}\}$, the relations in \eqref{eq:nonreference_pressure_reconstruction} produce the hydraulic state equivalent:
	\begin{equation}
		\boldsymbol x_{\rm H,pr}^{\rm B}
		=
		\boldsymbol Y_{\rm pr}
		\boldsymbol z_{\rm H}^{\rm B}
		+
		\widehat{\boldsymbol x}_{\rm H,pr}^{{\rm B},j},
		\label{eq:pressure_equivalent}
	\end{equation}
	where $\boldsymbol Y_{\rm pr} = \allowbreak \operatorname{col}\{ \allowbreak \boldsymbol S_{\rm pr,{\rm S}}^{\rm ref},\boldsymbol S_{\rm pr,{\rm R}}^{\rm ref}\}$ and $\widehat{\boldsymbol x}_{\rm H,pr}^{{\rm B},j}= \allowbreak \operatorname{col}\{ -\boldsymbol \ell_{\rm pr,{\rm S}}^{{\rm B},j}, \allowbreak \boldsymbol \ell_{\rm pr,{\rm R}}^{{\rm B},j}\}$. 
	The required supply-to-return pressure margin is then expressed in terms of the DHN boundary variables as follows:
	\begin{equation}
		(\boldsymbol S_{\rm pr,S}^{\rm ref}
		-\boldsymbol S_{\rm pr,R}^{\rm ref})
		\boldsymbol z_{\rm H}^{\rm B}
		\geq
		\boldsymbol \delta_{\rm pr}^{\rm B}
		+\boldsymbol \ell_{\rm pr,S}^{{\rm B},j}
		+\boldsymbol \ell_{\rm pr,R}^{{\rm B},j},
		\label{eq:projected_pressure_limits}
	\end{equation}
	where $\boldsymbol \delta_{\rm pr}^{\rm B}$ specifies the required pressure margin.
	
	\subsubsection{Projected Feasible Region of the DHN}
	Combining the thermal and hydraulic state equivalents in \eqref{eq:thermal_all_equivalent} and \eqref{eq:pressure_equivalent}, the projected temperature limits in \eqref{eq:projected_temperature_limits}, and the pressure margin constraint in \eqref{eq:projected_pressure_limits} yields the projected DHN feasible region:
	\begin{equation}
		\Omega_{\rm H}^{{\rm B}, j}
		= \left\{
		\boldsymbol z_{\rm H}^{\rm B}
		\ \middle|\
		\boldsymbol E_{\rm H}^j
		\boldsymbol z_{\rm H}^{\rm B}
		= \boldsymbol b_{\rm H}^j,
		\;
		\boldsymbol G_{\rm H}^j
		\boldsymbol z_{\rm H}^{\rm B}
		\leq \boldsymbol d_{\rm H}^j
		\right\},
		\label{eq:boundary_projection_compact}
	\end{equation}
	where $\boldsymbol E_{\rm H}^{j}$ and $\boldsymbol b_{\rm H}^{j}$ represent the equality relations derived from the thermal and hydraulic state equivalents; while the inequalities defined by $\boldsymbol G_{\rm H}^{j}$ and $\boldsymbol d_{\rm H}^{j}$ represent the projected temperature limits and the pressure margin constraint.
	
	\section{Decentralized Power Dispatch with Anderson Prediction}
	Under the VF-VT operation mode, the DHN subproblem yields an updated mass flow trajectory during the decentralized coordination, requiring the equivalent model to be updated at each iteration. To address this computational burden imposed by the iterative procedure, a prediction strategy is devised for the decentralized coordination.
	
	Section~IV-A presents the EPS and DHN subproblems and specifies the information that is exchanged between the two operators. Section~IV-B describes the iterative coordination scheme and the Anderson prediction procedure.
	
	\subsection{EPS and DHN Subproblem Formulations}
	At iteration $r$, the DHN operator constructs the projected feasible region $\Omega_{\rm H}^{{\rm B},(r)}$ of the DHN by setting $\boldsymbol x_{\rm m}^{{\rm B},j}=\boldsymbol x_{\rm m}^{{\rm B},(r)}$ in \eqref{eq:boundary_projection_compact}.
	The pump power function is evaluated at the same mass flow trajectory as follows:
	$\boldsymbol{\it\Pi}_{\rm pump}^{{\rm B},(r)}
	\left(\boldsymbol z_{\rm H}^{\rm B}\right)
	=
	\boldsymbol{\it\Pi}_{\rm pump}^{\rm B}
	\left(
	\boldsymbol x_{\rm m}^{{\rm B},(r)},
	\boldsymbol z_{\rm H}^{\rm B}
	\right).$
	The resulting coefficients $\boldsymbol E_{\rm H}^{(r)}$, $\boldsymbol b_{\rm H}^{(r)}$, $\boldsymbol G_{\rm H}^{(r)}$, and $\boldsymbol d_{\rm H}^{(r)}$, together with ${\boldsymbol{\it \Pi}}_{\rm pump}^{{\rm B},(r)}(\cdot)$, are transmitted to the EPS operator, while the internal DHN variables remain local to the DHN operator.
	
	The decentralized procedure comprises two stages. Stage I alternates between the EPS feasibility checking problem and the DHN subproblem to obtain a feasible initial solution. Stage II alternates between the EPS dispatch problem and the DHN subproblem to minimize the operating cost of the IHPS.
	
	In Stage I, the EPS operator solves the feasibility checking problem. The nonnegative slack variables $\boldsymbol{\sigma}_{\rm eq}\geq\boldsymbol 0$ and $\boldsymbol{\sigma}_{\rm ineq}\geq\boldsymbol 0$ relax the equality and inequality relations defining $\Omega_{\rm H}^{{\rm B},(r)}$, yielding the following problem:
	\begin{align}
		\mathrm{P}_{\rm E,F}^{(r)}\!:\hspace{2.5em}
		& \min_{\makebox[0pt][c]{$\scriptstyle
				\boldsymbol{x}_{\rm E}^{\rm B},
				\boldsymbol{z}_{\rm E}^{\rm B},
				\boldsymbol{z}_{\rm H}^{\rm B},
				\boldsymbol{\sigma}_{\rm eq},
				\boldsymbol{\sigma}_{\rm ineq}$}}
		\hspace{3.5em}
		\boldsymbol{1}^{\top}\boldsymbol{\sigma}_{\rm eq}
		+
		\boldsymbol{1}^{\top}\boldsymbol{\sigma}_{\rm ineq}
		\label{eq:PEF_obj}
		\\
		& \mathrm{s.t.}\quad
		\eqref{eq:P0_elec},\ \eqref{eq:P0_chp},
		\notag
		\\
		& \phantom{\mathrm{s.t.}\quad}
		\boldsymbol{C}_{\rm pump,z}^{\rm B}
		\boldsymbol{z}_{\rm E}^{\rm B}
		=
		\boldsymbol {\it \Pi}_{\rm pump}^{{\rm B},(r)}
		(\boldsymbol{z}_{\rm H}^{\rm B}),
		\label{eq:PEF_pump}
		\\
		& \phantom{\mathrm{s.t.}\quad}
		-\boldsymbol{\sigma}_{\rm eq}
		\leq
		\boldsymbol{E}_{\rm H}^{(r)}
		\boldsymbol{z}_{\rm H}^{\rm B}
		-
		\boldsymbol{b}_{\rm H}^{(r)}
		\leq
		\boldsymbol{\sigma}_{\rm eq},
		\label{eq:PEF_eq_slack}
		\\
		& \phantom{\mathrm{s.t.}\quad}
		\boldsymbol{G}_{\rm H}^{(r)}
		\boldsymbol{z}_{\rm H}^{\rm B}
		\leq
		\boldsymbol{d}_{\rm H}^{(r)}
		+
		\boldsymbol{\sigma}_{\rm ineq}.
		\label{eq:PEF_ineq_slack}
	\end{align}
	
	Stage I terminates if $\max \{ \| \boldsymbol{\sigma}_{\rm eq}^{(r)} \|_{\infty}, \| \boldsymbol{\sigma}_{\rm ineq}^{(r)} \|_{\infty} \} \leq \epsilon_{\rm F}$, where $\epsilon_{\rm F}>0$ is a predefined tolerance level. The resulting boundary variables and mass flow trajectory then initialize Stage II. If this condition is not satisfied, the DHN subproblem $\mathrm P_{\rm H}^{(r)}$ is solved, and Stage~I proceeds to the next iteration.
	
	In Stage II, the EPS operator minimizes the operating cost of the IHPS by solving the following problem:
	\begin{align}
		\mathrm{P}_{\rm E,O}^{(r)}:\!
		& \min_{
			\boldsymbol{x}_{\rm E}^{\rm B},
			\boldsymbol{z}_{\rm E}^{\rm B},
			\boldsymbol{z}_{\rm H}^{\rm B}}
		\quad
		J_{\rm op}^{\rm B}
		(
		\boldsymbol{x}_{\rm E}^{\rm B},
		\boldsymbol{z}_{\rm E}^{\rm B},
		\boldsymbol{z}_{\rm H}^{\rm B}
		)
		\label{eq:PEO_obj}
		\\
		& \mathrm{s.t.}\quad
		\eqref{eq:P0_elec},\ \eqref{eq:P0_chp},
		\notag
		\\
		& \phantom{\mathrm{s.t.}\quad}
		\boldsymbol{E}_{\rm H}^{(r)}
		\boldsymbol{z}_{\rm H}^{\rm B}
		=
		\boldsymbol{b}_{\rm H}^{(r)},
		\;
		\boldsymbol{G}_{\rm H}^{(r)}
		\boldsymbol{z}_{\rm H}^{\rm B}
		\leq
		\boldsymbol{d}_{\rm H}^{(r)},
		\label{eq:PEO_projection}
		\\
		& \phantom{\mathrm{s.t.}\quad}
		\boldsymbol{C}_{\rm pump,z}^{\rm B}
		\boldsymbol{z}_{\rm E}^{\rm B}
		=
		\boldsymbol {\it \Pi}_{\rm pump}^{{\rm B},(r)}
		(\boldsymbol{z}_{\rm H}^{\rm B}).
		\label{eq:PEO_pump}
	\end{align}
	
	The resulting boundary variables $\boldsymbol z_{\rm E}^{{\rm B},(r)}$ and $\boldsymbol z_{\rm H}^{{\rm B},(r)}$ are transmitted to the DHN operator.
	
	The DHN subproblem is solved in both stages. Upon receiving $\boldsymbol z_{\rm E}^{{\rm B},(r)}$ and $\boldsymbol z_{\rm H}^{{\rm B},(r)}$ from the EPS operator, the DHN operator solves the following subproblem:
	\begin{align}
		\mathrm{P}_{\rm H}^{(r)}:\!
		& \min_{
			\boldsymbol{x}_{\rm H,T}^{\rm B},
			\boldsymbol{x}_{\rm H,pr}^{\rm B},
			\boldsymbol{x}_{\rm m}^{\rm B},
			\boldsymbol{z}_{\rm H}^{\rm B}}
		\quad
		\left\|
		\boldsymbol{z}_{\rm H}^{\rm B}
		- \boldsymbol{z}_{\rm H}^{{\rm B},(r)}
		\right\|_2^2
		\label{eq:PH_obj}
		\\
		& \mathrm{s.t.}\quad
		\eqref{eq:P0_heat_alg}, \,
		\eqref{eq:P0_mass} \, \text{--} \, \eqref{eq:P0_pressure}, \,
		\eqref{eq:pipe_augmented_relation},
		\notag
		\\
		& \phantom{\mathrm{s.t.}\quad}
		\boldsymbol{\it \Gamma}_{\rm CHP}^{\rm B}
		( \boldsymbol z_{\rm E}^{{\rm B},(r)},
		\boldsymbol z_{\rm H}^{\rm B} )
		\leq \boldsymbol 0,
		\label{eq:PH_chp}
		\\
		& \phantom{\mathrm{s.t.}\quad}
		\boldsymbol{C}_{\rm pump,z}^{\rm B}
		\boldsymbol z_{\rm E}^{{\rm B},(r)}
		=
		\boldsymbol{\it \Pi}_{\rm pump}^{\rm B}
		( \boldsymbol x_{\rm m}^{\rm B},
		\boldsymbol z_{\rm H}^{\rm B} ).
		\label{eq:PH_pump}
	\end{align}
	
	Since $\mathrm P_{\rm H}^{(r)}$ retains the Bernstein--Galerkin thermal relations underlying $\Omega_{\rm H}^{{\rm B},(r)}$, the updated mass flow trajectory $\boldsymbol x_{\rm m}^{{\rm B},(r+1)}$ is used to update the DHN equivalent model for the next iteration within the same thermal formulation.
	
	\subsection{EPS--DHN Coordination With Anderson Prediction}
	\subsubsection{Alternating EPS--DHN Coordination}
	Starting from the feasible initial point obtained in Stage I, Stage II alternates between solving the EPS dispatch problem $\mathrm P_{\rm E,O}^{(r)}$ and the DHN subproblem $\mathrm P_{\rm H}^{(r)}$. At iteration $r$, the projected feasible region derived for the DHN in Section III is evaluated at $\boldsymbol x_{\rm m}^{{\rm B},(r)}$ and embedded in $\mathrm P_{\rm E,O}^{(r)}$. The EPS operator transmits the resulting boundary variables to the DHN operator, which solves $\mathrm P_{\rm H}^{(r)}$ and obtains $\boldsymbol x_{\rm m}^{{\rm B},(r+1)}$ for reconstructing the DHN equivalent at the next iteration. The successive solutions of $\mathrm P_{\rm E,O}^{(r)}$ and $\mathrm P_{\rm H}^{(r)}$ constitute one alternating EPS--DHN coordination update.
	
	Accordingly, the alternating EPS--DHN coordination scheme can be expressed as a fixed-point iteration on the coordination vector $\boldsymbol{\varpi}^{(r)} = \operatorname{col}\{\boldsymbol{x}_{\rm m}^{{\rm B},(r)}, \boldsymbol{z}_{\rm H}^{{\rm B},(r)}\}$, with the alternating update and fixed-point residual represented as follows:
	\begin{equation}
		\begin{aligned}
			&\boldsymbol{\varpi}_{\rm alt}^{(r+1)}
			= {\it\Phi}
			( \boldsymbol{\varpi}^{(r)} ), \; \boldsymbol{\zeta}^{(r)}
			= \boldsymbol{\varpi}_{\rm alt}^{(r+1)}
			- \boldsymbol{\varpi}^{(r)},
		\end{aligned}
		\label{eq:fixed_point_map}
	\end{equation}
	where $\it \Phi(\cdot)$ represents the alternating update obtained by reconstructing the DHN equivalent and subsequently solving $\mathrm P_{\rm E,O}^{(r)}$ and $\mathrm P_{\rm H}^{(r)}$, and $\boldsymbol\zeta^{(r)}$ represents the mismatch between the current coordination vector and its alternating update.
	
	During the naive alternating iteration, the coordination vector of the next iteration is set directly to its latest update, so the alternating coordination scheme follows the direct fixed-point update defined by $\it\Phi(\cdot)$. Such alternating updates are conducted iteratively, each involving the reconstruction of the flow-dependent DHN equivalent and the solving of $\mathrm P_{\rm E,O}^{(r)}$ and $\mathrm P_{\rm H}^{(r)}$. This procedure is inefficient, particularly when the fixed-point iteration converges slowly. To achieve improved efficiency, the Anderson prediction strategy is devised to exploit the historic alternating updates and fixed-point residuals to construct the subsequent coordination vector.
	
	To account for the different numerical scales of the coordination variables, the residual is scaled as follows:
	\begin{equation}
		\boldsymbol{\zeta}_{\rm sc}^{(r)}
		= \boldsymbol W_{\varpi}
		\boldsymbol{\zeta}^{(r)},
		\;
		\boldsymbol W_{\varpi}
		= \operatorname{diag}\{s_{\rm m}^{-1}\boldsymbol I_{n_{\rm m}},s_{\rm H}^{-1}\boldsymbol I_{n_{\rm H}}\},
		\label{eq:scaled_residual}
	\end{equation}
	where $n_{\rm m}$ and $n_{\rm H}$ denote the dimensions of $\boldsymbol x_{\rm m}^{\rm B}$ and $\boldsymbol z_{\rm H}^{\rm B}$, respectively. The scaling factors $s_{\rm m}=\max\{1,\|\boldsymbol x_{\rm m}^{\rm B,0}\|_{\infty}\}$ and $s_{\rm H}=\max\{1,\|\boldsymbol z_{\rm H}^{\rm B,0}\|_{\infty}\}$ are set upon the initialization of Stage II and remain fixed in the subsequent iterations. The scaled residuals and the alternating updates derived from the recent iterations are retained for constructing the Anderson prediction scheme described below.
	
	\subsubsection{Safeguarded Anderson Prediction}
	At iteration $r$, the Anderson memory depth is $n_{\rm A}^{(r)}=\min\{n_{\rm A}^{\max},r\}$, and the weight vector $\boldsymbol\kappa^{(r)}=\operatorname{col}\{\kappa_0^{(r)},\ldots,\kappa_{n_{\rm A}}^{(r)}\}$ is obtained by solving the following regularized least-squares problem:
	\begin{align}
		\boldsymbol{\kappa}^{(r)}
		\in \arg\min_{\boldsymbol\kappa} \;
		& \big\|
		\sum\nolimits_{i=0}^{n_{\rm A}^{(r)}}
		\kappa_i
		\boldsymbol{\zeta}_{\rm sc}^{
			r-n_{\rm A}^{(r)}+i}
		\big\|_2^2
		+
		\mu_{\rm A}\|\boldsymbol\kappa\|_2^2
		\label{eq:anderson_ls}
		\\
		\mathrm{s.t.} \;
		& \sum\nolimits_{i=0}^{n_{\rm A}^{(r)}}
		\kappa_i=1,
		\;
		\|\boldsymbol\kappa\|_1\leq\kappa_{\max},
		\label{eq:anderson_affine}
	\end{align}
	where $n_{\rm A}^{\max}$ is a predefined maximum memory depth, $\mu_{\rm A}>0$ regularizes the least-squares problem, and $\kappa_{\max}$ specifies the $\ell_1$-norm bound that is imposed on the Anderson weights.
	
	The resulting weights form the Anderson prediction $\boldsymbol{\varpi}_{\rm A}^{(r+1)}$ from the historic alternating updates, with its scaled displacement from the latest update $\boldsymbol{\varpi}_{\rm alt}^{(r+1)}$ denoted by $\boldsymbol d_{\rm A}^{(r)}$:
	\begin{equation}
		\boldsymbol{\varpi}_{\rm A}^{(r+1)}
		= \sum\nolimits_{i=0}^{n_{\rm A}^{(r)}}
		\kappa_i^{(r)}
		\boldsymbol{\varpi}_{\rm alt}^{
			r-n_{\rm A}^{(r)}+i+1},
		\label{eq:anderson_prediction}
	\end{equation}
	\begin{equation}
		\boldsymbol d_{\rm A}^{(r)}
		= \boldsymbol W_{\varpi}
		( \boldsymbol{\varpi}_{\rm A}^{(r+1)}
		- \boldsymbol{\varpi}_{\rm alt}^{(r+1)} ).
		\label{eq:anderson_displacement}
	\end{equation}
	
	Given a prescribed displacement limit $\delta_{\max}>0$, the safeguarded prediction $\boldsymbol{\varpi}_{\rm pred}^{(r+1)}$ and the associated scaling factor $\alpha^{(r)}$ are defined as follows:
	\begin{align}
		\boldsymbol{\varpi}_{\rm pred}^{(r+1)} 
		& = \boldsymbol{\varpi}_{\rm alt}^{(r+1)} 
		+ \alpha^{(r)} ( \boldsymbol{\varpi}_{\rm A}^{(r+1)} 
		- \boldsymbol{\varpi}_{\rm alt}^{(r+1)} ),
		\label{eq:safeguarded_prediction}
		\\
		\alpha^{(r)}
		& = \begin{cases}
			1,
			& \|\boldsymbol{d}_{\rm A}^{(r)}\|_{\infty} \leq \delta_{\max}, 
			\\
			\delta_{\max} / \|\boldsymbol{d}_{\rm A}^{(r)}\|_{\infty}, 
			& \|\boldsymbol{d}_{\rm A}^{(r)}\|_{\infty} > \delta_{\max},
		\end{cases}
		\label{eq:prediction_factor}
	\end{align}
	
	Consequently, the safeguard bounds the scaled displacement of the predicted coordination vector from the latest update:
	\begin{equation}
		\| \boldsymbol W_{\varpi}
		( \boldsymbol{\varpi}_{\rm pred}^{(r+1)}
		- \boldsymbol{\varpi}_{\rm alt}^{(r+1)} ) \|_{\infty}
		\leq \delta_{\max}.
		\label{eq:safeguard_bound}
	\end{equation}
	
	\subsubsection{Termination Criterion and Iteration Restart Strategy}
	The safeguarded prediction $\boldsymbol{\varpi}_{\rm pred}^{(r+1)}$ is used as the coordination vector for iteration $r+1$. No additional EPS or DHN subproblem is solved at $\boldsymbol{\varpi}_{\rm pred}^{(r+1)}$ during iteration $r$. The alternating update $\boldsymbol{\varpi}_{\rm alt}^{(r+1)}$ is retained for subsequent prediction.
	Let $J_{\rm alt}^{(r+1)}$ denote the operating cost obtained from $\mathrm P_{\rm E,O}^{(r)}$ during the alternating update. The relative cost error is evaluated as follows:
	\begin{equation}
		\operatorname{err}_{J}^{(r+1)}
		= | J_{\mathrm{alt}}^{(r+1)} - J_{\mathrm{alt}}^{(r)} | /
		\max \bigl\{ 1, | J_{\mathrm{alt}}^{(r+1)} | \bigr\}.
		\label{eq:cost_convergence}
	\end{equation}
	
	Stage II terminates when $\operatorname{err}_{J}^{(r+1)}\leq\epsilon_{\rm J}$, where $\epsilon_{\rm J}>0$ is a predefined tolerance, and the latest successful alternating EPS--DHN solution is retained as the final dispatch result. Otherwise, $\boldsymbol{\varpi}_{\rm pred}^{(r+1)}$ is used as the coordination vector for iteration $r+1$, while $\boldsymbol{\varpi}_{\rm alt}^{(r+1)}$ is retained as the restart point. If either $\mathrm P_{\rm E,O}^{(r+1)}$ or $\mathrm P_{\rm H}^{(r+1)}$ fails under the predicted vector, $\boldsymbol{\varpi}^{(r+1)}$ is reset to $\boldsymbol{\varpi}_{\rm alt}^{(r+1)}$, and iteration $r+1$ is restarted from $\boldsymbol{\varpi}_{\rm alt}^{(r+1)}$.
	
	The decentralized coordination procedure is summarized in Algorithm~\ref{alg:distributed_dispatch}. Stage I provides a feasible initial point for Stage II. In Stage II, the safeguarded Anderson prediction determines the next coordination vector, while the latest alternating update is retained as the restart point if either subproblem fails under the predicted vector. This preserves the alternating EPS--DHN procedure without additional subsystem solving steps.
	
	\begingroup
	\setlength{\intextsep}{3pt plus 1pt minus 1pt}
	\begin{algorithm}[H]
		\caption{Decentralized IHPS Power Dispatch with Anderson Prediction}
		\label{alg:distributed_dispatch}
		\footnotesize
		\begin{algorithmic}[1]
			\State \textbf{Initialization:} set $r=0$ and initialize
			$\boldsymbol x_{\rm m}^{\rm B,0}$.
			
			\State \textbf{Stage I: Feasible Initialization.}
			\Repeat
			\State Evaluate $\Omega_{\rm H}^{{\rm B},(r)}$ and $ \boldsymbol{\it \Pi}_{\rm pump}^{{\rm B},(r)}(\cdot)$ at $\boldsymbol x_{\rm m}^{{\rm B},(r)}$.
			
			\State Solve $\mathrm P_{\rm E,F}^{(r)}$ in \eqref{eq:PEF_obj}--\eqref{eq:PEF_ineq_slack}, obtaining $( \boldsymbol z_{\rm E}^{{\rm B},(r)}, \boldsymbol z_{\rm H}^{{\rm B},(r)}, \boldsymbol\sigma_{\rm eq}^{(r)}, \boldsymbol\sigma_{\rm ineq}^{(r)} )$.
			
			\If{ $\max\{ \|\boldsymbol\sigma_{\rm eq}^{(r)}\|_\infty,
				\|\boldsymbol\sigma_{\rm ineq}^{(r)}\|_\infty \}>\epsilon_{\rm F}$ }
			
			\State Solve $\mathrm P_{\rm H}^{(r)}$ in \eqref{eq:PH_obj}--\eqref{eq:PH_pump}, update $\boldsymbol x_{\rm m}^{{\rm B},(r+1)}$, set $r\leftarrow r+1$.
			\EndIf
			
			\Until{ $\max\{ \|\boldsymbol\sigma_{\rm eq}^{(r)}\|_\infty,
				\|\boldsymbol\sigma_{\rm ineq}^{(r)}\|_\infty \} \leq \epsilon_{\rm F}$ }.
			
			\State \textbf{Stage II: Operating Cost Minimization.}
			\State Set $r\leftarrow0$, initialize
			$(\boldsymbol\varpi^{0},J_{\rm alt}^{0})$ from Stage I.
			
			\Repeat
			\State Evaluate $\Omega_{\rm H}^{{\rm B},(r)}$ and
			$\boldsymbol{\it \Pi}_{\rm pump}^{{\rm B},(r)}(\cdot)$
			at $\boldsymbol x_{\rm m}^{{\rm B},(r)}$.
			
			\State Solve $\mathrm P_{\rm E,O}^{(r)}$ in
			\eqref{eq:PEO_obj}--\eqref{eq:PEO_pump}, obtaining
			$(\boldsymbol z_{\rm E}^{{\rm B},(r)},
			\boldsymbol z_{\rm H}^{{\rm B},(r)},
			J_{\rm alt}^{(r+1)})$.
			
			\If{$\mathrm P_{\rm E,O}^{(r)}$ fails under
				$\boldsymbol\varpi_{\rm pred}^{(r)}$}
			\State Set
			$(\boldsymbol\varpi^{(r)},n_{\rm A}^{(r)})
			\leftarrow
			(\boldsymbol\varpi_{\rm alt}^{(r)},0)$
			and restart iteration $r$.
			\EndIf
			
			\State Compute $\operatorname{err}_{J}^{(r+1)}$ using
			\eqref{eq:cost_convergence}.
			
			\If{$\operatorname{err}_{J}^{(r+1)}>\epsilon_{\rm J}$}
			\State Solve $\mathrm P_{\rm H}^{(r)}$ in
			\eqref{eq:PH_obj}--\eqref{eq:PH_pump};
			update $\boldsymbol x_{\rm m}^{{\rm B},(r+1)}$.
			
			\If{$\mathrm P_{\rm H}^{(r)}$ fails under
				$\boldsymbol\varpi_{\rm pred}^{(r)}$}
			\State Set
			$(\boldsymbol\varpi^{(r)},n_{\rm A}^{(r)})
			\leftarrow
			(\boldsymbol\varpi_{\rm alt}^{(r)},0)$
			and restart iteration $r$.
			\EndIf
			
			\State Compute $\boldsymbol\varpi_{\rm alt}^{(r+1)}$,
			$\boldsymbol\zeta^{(r)}$, and
			$\boldsymbol\zeta_{\rm sc}^{(r)}$ 
			using
			\eqref{eq:fixed_point_map} and
			\eqref{eq:scaled_residual}.
			
			\State Set
			$\boldsymbol\varpi^{(r+1)}
			\leftarrow
			\boldsymbol\varpi_{\rm alt}^{(r+1)}$.
			
			\If{$n_{\rm A}^{(r)}\geq1$}
			\State Compute
			$\boldsymbol\varpi_{\rm pred}^{(r+1)}$ using
			\eqref{eq:anderson_ls}--\eqref{eq:prediction_factor},
			set $\boldsymbol\varpi^{(r+1)}
			\leftarrow \boldsymbol\varpi_{\rm pred}^{(r+1)}$.
			\EndIf
			\EndIf
			
			\State Set $r\leftarrow r+1$.
			\Until{$\operatorname{err}_{J}^{(r)}\leq\epsilon_{\rm J}$}
			
			\State \textbf{Output:} Final IHPS power dispatch and operating cost.
			
		\end{algorithmic}
		\normalsize
	\end{algorithm}
	\endgroup
	
	The above coordination procedure is designed for the VF-VT operation mode. The proposed method also applies to the CF-VT and the variable-flow and constant-temperature (VF-CT) operation modes. Under the CF-VT operation mode, the mass flow trajectories are prescribed, and the heat load demand is supplied through temperature regulation. The DHN model then becomes linear, requiring only a single DHN equivalent projection and degenerating the proposed method to a noniterative decentralized paradigm. Under the VF-CT operation mode, the supply temperature of the heat source is prescribed, while the mass flow trajectories and the other thermal states remain variable. Therefore, the DHN model remains nonlinear, and the DHN equivalent is updated with the mass flow during the iterative EPS--DHN coordination.
	
	\section{Case Studies}
	The proposed method is evaluated using two IHPSs of different scales. A small-scale system, comprising a six-bus EPS and a six-node DHN, namely, the E6-H6 IHPS, is used in Section V-A for a detailed evaluation of the achieved decentralized dispatch performance. A large-scale system, comprising a modified IEEE 39-bus EPS and a 20-node DHN, namely, the E39-H20 IHPS, is used in Section V-B to evaluate scalability. The proposed decentralized Bernstein--Galerkin equivalent projection method (BGEP-D) is benchmarked against the centralized Bernstein--Galerkin method (BGM-C) \cite{deng2026integrated}, the decentralized node method-based sequential equivalent (NMSE-D) \cite{zheng2023distributed}, and the centralized node method (NM-C) \cite{li2015ComebinedHeat}. The detailed data of the two IHPSs are provided in \cite{deng2026supplementary}.
	
	The Bernstein degrees are set to $n_t=n_s=3$. The maximum numbers of iterations for Stages I and II are set to 50 and 200, respectively. The Anderson parameters are $n_{\rm A}^{\max}=3$, $\mu_{\rm A}=\sciresult{1}{-6}$, $\kappa_{\max}=5$, and $\delta_{\max}=0.20$, with tolerances of $\epsilon_{\rm F}=\sciresult{1}{-4}$, $\epsilon_{\rm J}=\sciresult{1}{-4}$. All tests are implemented in MATLAB R2023a on a computer with 16 GB of RAM and an Intel Core i5-13400 CPU.
	
	\subsection{Simulation on a Small-Scale IHPS}
	\begingroup
	\setlength{\intextsep}{3pt plus 1pt minus 1pt}
	\setlength{\floatsep}{3pt plus 1pt minus 1pt}
	\setlength{\textfloatsep}{3pt plus 1pt minus 1pt}
	\setlength{\abovecaptionskip}{3pt}
	\setlength{\belowcaptionskip}{0.5pt}
	The decentralized coordination scheme is evaluated on the E6-H6 IHPS under the VF-VT operation mode, with mass flow and temperature trajectories updated at each iteration.
	
	\standalonesubsubsection{Economic Performance and Coordinated Operation}
	The economic and flexibility performance of the tested methods is summarized in Table~\ref{tab:caseA_operating_performance}. The proposed BGEP-D attains the lowest operating cost, 1.47\% lower than that of BGM-C, while accommodating all available wind power generation. Under the VF-VT operation mode, the joint optimization of mass flow rates and temperatures renders the DHN subproblem nonconvex, and the decentralized and centralized procedures yield different feasible operating points.
	
	The comparison between the two decentralized methods highlights the advantages of BGEP-D. Compared with NMSE-D, BGEP-D reduces the total cost by \$1,798.6 and eliminates 7.48 MWh of wind curtailment. Moreover, both Bernstein--Galerkin cases achieve full wind accommodation, whereas the node method-based cases incur 7.48--18.61 MWh of wind curtailment. These differences arise from different DHN equivalents supplied to the EPS. BGEP-D constructs the DHN equivalent in the Bernstein space and retains continuous spatiotemporal thermal dynamics under the updated mass flow trajectory. In contrast, NMSE-D represents thermal dynamics through discrete-time node method-based relations and does not explicitly represent thermal variations within each interval.

	\begin{table}[H]
		\caption{Economic and Flexibility Performance on the E6-H6 IHPS}
		\label{tab:caseA_operating_performance}
		\centering
		\footnotesize
		\renewcommand{\arraystretch}{1}
		\begin{tabular*}{\columnwidth}{@{\extracolsep{\fill}}ccccc@{}}
			\toprule
			Method & Total & Gen. Cost & Curtailment Cost & Wind Curtailment \\
			& (\$)       & (\$)      & (\$)      & (MWh)      \\
			\midrule
			BGEP-D & 25,684.7 & 25,684.7 & 0.00   & 0.00  \\
			BGM-C  & 26,068.7 & 26,068.7 & 0.00   & 0.00  \\
			NMSE-D & 27,483.3 & 27,175.5 & 307.7  & 7.48  \\
			NM-C   & 26,838.9 & 26,073.6 & 765.3  & 18.61 \\
			\bottomrule
		\end{tabular*}
		\vspace{-0.4em}
	\end{table}
    
	The resulting operation schedules are further shown in Fig.~\ref{fig1:caseA_operating_summary}. During the peak power demand intervals (9--15 h), compared with NMSE-D, BGEP-D results in a higher CHP power output and a greater CHP mass flow. The greater CHP mass flow accelerates the heat transport, allowing a higher CHP power output to be obtained while meeting the heat demand. This coordination exploits the electrothermal flexibility of the CHP unit, yielding the lowest operating cost and full wind accommodation. In contrast, NMSE-D results in a lower CHP power output during the peak period, whereas its CHP mass flow increases after 15 h. The mismatch between the CHP power output and DHN heat transport in the NMSE-D case limits the CHP utilization, increases the operating cost and results in wind curtailment.
	
	\begin{figure}[H]
		\centering
		\includegraphics[width=0.9\columnwidth]{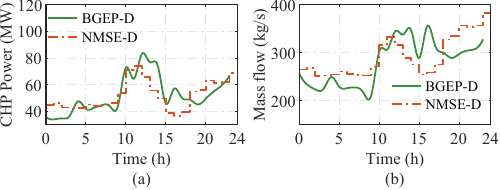}
		\vspace{-0.5em}
		\caption{Comparison of operation schedules in the E6-H6 IHPS under the VF-VT operation mode: (a) CHP power output and (b) CHP mass flow.}
		\label{fig1:caseA_operating_summary}
		\vspace{-0.3em}
	\end{figure}
	
	\standalonesubsubsection{Consistency of DHN Thermal Representations}
	At each successful Stage II iteration, the internal thermal states obtained from the DHN subproblem are compared with those reconstructed from the DHN equivalent under the same $\boldsymbol x_{\rm m}^{{\rm B},(r)}$ and $\boldsymbol z_{\rm H}^{{\rm B},(r)}$. To evaluate the consistency of the DHN thermal dynamics throughout the decentralized coordination, the average and maximum closure gaps are defined as follows:
	\begin{align}
		&\bar{\eta}_{\mathrm{cl}}^{(r)}
		= \big\| 
		\boldsymbol x_{\mathrm{H,T}}^{{\rm B},(r)}
		- (\boldsymbol Y_{\mathrm{H}}^{(r)}
		\boldsymbol z_{\mathrm{H}}^{{\rm B},(r)}
		+ \widehat{\boldsymbol x}_{\mathrm{H,T}}^{{\rm B},(r)} )
		\big\|_{1}
		/ \operatorname{dim}
		(\boldsymbol x_{\mathrm{H,T}}^{{\rm B},(r)}), 
		\nonumber
		\\
		& \quad \eta_{\mathrm{cl},\infty}^{(r)}
		=\big\| 
		\boldsymbol x_{\mathrm{H,T}}^{{\rm B},(r)}
		- ( \boldsymbol Y_{\mathrm{H}}^{(r)}
		\boldsymbol z_{\mathrm{H}}^{{\rm B},(r)}
		+ \widehat{\boldsymbol x}_{\mathrm{H,T}}^{{\rm B},(r)} )
		\big\|_{\infty}.
		\label{eq:projection_update_closure}
	\end{align}
	
	The componentwise maximum gaps for the source return, load return, pipeline outlet, and nodal mixing temperatures are further evaluated to locate specific thermal discrepancies.
	
	Fig.~\ref{fig2:caseA_model_closure}(a) shows that the average closure gap of BGEP-D remains zero throughout the Stage-II iterations. For NMSE-D, the average closure gap reaches 3.53 $^\circ{\rm C}$, whereas the componentwise maximum gaps reach approximately 16 $^\circ{\rm C}$, as shown in Fig.~\ref{fig2:caseA_model_closure}(b). These gaps arise from the thermal formulations used in the DHN equivalent and subproblem. BGEP-D derives both formulations from the same Bernstein--Galerkin thermal relations. In contrast, NMSE-D derives the DHN equivalent from the discrete-time node method, whereas its DHN subproblem uses a surrogate thermal model. Accordingly, NMSE-D yields different internal thermal states under the same mass flow trajectory and boundary variable.
	
	As shown in Fig.~\ref{fig2:caseA_model_closure}(b), the closure gaps of NMSE-D persist across all four temperature components, with the pipeline outlet and nodal mixing temperatures exhibiting the largest values. These persistent gaps indicate inconsistent internal thermal state reconstructions between the DHN equivalent and subproblem of NMSE-D throughout Stage II. In contrast, the zero closure gap of BGEP-D shows that its two thermal representations remain consistent as the mass flow trajectory is updated.
	
	\begin{figure}[H]
		\centering
		\includegraphics[width=0.9\columnwidth]{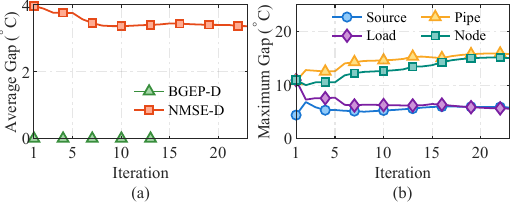}
		\vspace{-0.5em}
		\caption{Consistency of the DHN thermal representations in the E6-H6 IHPS: (a) average closure gaps over the Stage-II iterations and (b) componentwise maximum closure gaps of NMSE-D.}
		\label{fig2:caseA_model_closure}
		\vspace{-0.3em}
	\end{figure}
	
	\standalonesubsubsection{Computational Performance}
	The computational performance attained by the four methods is summarized in Table~\ref{tab:caseA_runtime}. BGEP-D completes Stages I and II in 2 and 14 iterations, respectively, whereas NMSE-D requires 3 and 23 iterations, respectively. Their total runtimes are 8.43~s and 16.57~s, respectively. Among the centralized methods, BGM-C and NM-C require 6.78~s and 42.88~s, respectively.
	\vspace{-0.3em}
	
	\begin{table}[H]
		\caption{Computational Performance on the E6-H6 IHPS}
		\label{tab:caseA_runtime}
		\centering
		\footnotesize
		\begin{tabular*}{\columnwidth}{@{\extracolsep{\fill}} l c c c c @{}}
			\toprule
			\multirow{2}{*}[-0.5ex]{Method}
			& \multirow{2}{*}[-0.5ex]{\makecell[c]{Iterations\\(Stages I$+$II)}}
			& \multicolumn{3}{c}{Runtime (s)} \\
			\cmidrule(lr){3-5}
			& & Optimization & Prediction & Total \\
			\midrule
			BGEP-D & 2$+$14 & 8.42  & 0.01 & 8.43  \\
			BGM-C  & --       & 6.78  & --   & 6.78  \\
			NMSE-D & 3$+$23 & 16.57 & --   & 16.57 \\
			NM-C   & --       & 42.88 & --   & 42.88 \\
			\bottomrule
		\end{tabular*}
	\end{table}
	The methods based on the Bernstein--Galerkin formulation exhibit lower runtimes than the node method benchmarks do under both decentralized and centralized implementations. With respect to the centralized methods, BGM-C is solved without internal iterations, whereas NM-C requires internal iterations to update the flow-dependent thermal model. Compared with BGM-C, BGEP-D requires only 1.65~s more runtime despite the iterative EPS--DHN coordination, while reducing the runtime by 8.14~s compared with NMSE-D. For the decentralized methods, this runtime reduction is further associated with fewer coordination iterations. The Anderson prediction process accounts for only 0.01~s of the total runtime and thus introduces negligible computational overhead.
	
	Fig.~\ref{fig3:caseA_convergence} compares the Stage-II convergence performances of BGEP-D and NMSE-D in terms of operating cost and relative cost error. The operating cost incurred by BGEP-D decreases rapidly during the early-stage iterations, and the termination criterion is satisfied in 14 iterations. In contrast, the operating cost of NMSE-D remains relatively high, and the termination criterion is met in 23 iterations, with dramatic fluctuations in the relative cost error. Together with the closure gaps shown in Fig.~\ref{fig2:caseA_model_closure}, the convergence results suggest that the thermal state inconsistency between the DHN equivalent and the DHN subproblem of NMSE-D contributes to iterative coordination adjustments before reaching convergence. These results indicate that BGEP-D provides improved computational performance while maintaining thermal consistency between the DHN equivalent and subproblem.
	
	\begin{figure}[H]
		\centering
		\includegraphics[width=0.9\columnwidth]{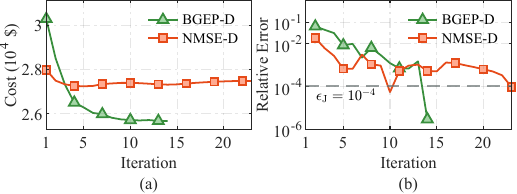}
		\vspace{-0.5em}
		\caption{Stage-II convergence trends of BGEP-D and NMSE-D on the E6-H6 IHPS: (a) operating costs and (b) relative cost errors.}
		\label{fig3:caseA_convergence}
		\vspace{-0.3em}
	\end{figure}
	
	\standalonesubsubsection{Effect of the Anderson Prediction}
	The effect of the Anderson prediction is studied by comparing BGEP-D variants with and without Anderson prediction. As reported in Table~\ref{tab:caseA_pc_ablation}, incorporating Anderson prediction reduces the number of Stage-II iterations from 31 to 14 and the total runtime from 15.23~s to 8.43~s while yielding a lower operating cost. Both cases satisfy the termination tolerance.
	
	\vspace{-0.3em}
	\begin{table}[H]
		\caption{Comparison of Performance With and Without Anderson Prediction on the E6-H6 IHPS}
		\label{tab:caseA_pc_ablation}
		\centering
		\footnotesize
		\setlength{\tabcolsep}{2.0pt}
		\renewcommand{\arraystretch}{1.04}
		\begin{tabular*}{\columnwidth}{@{\extracolsep{\fill}}ccccc@{}}
			\toprule
			Method
			& \makecell[c]{Iterations\\(Stage II)}
			& \makecell[c]{Operating Cost\\(\$)}
			& Final Cost Error
			& \makecell[c]{Runtime\\(s)} 
			\\
			\midrule
			BGEP-D & 14 & 25,684.69 & \sciresult{2.86}{-6} & 8.43 
			\\
			BGEP-D (No Pred.) & 31 & 25,952.94 & \sciresult{8.73}{-5} & 15.23 
			\\
			\bottomrule
		\end{tabular*}
	\end{table}
	
	Fig.~\ref{fig4:caseA_pc_ablation} further compares the corresponding Stage-II convergence trends. With Anderson prediction, the operating cost approaches its final level with fewer coordination updates, and the relative cost error reaches the termination threshold earlier. The prediction method is used to construct the coordination vector for the next iteration from recent alternating updates and fixed-point residuals, thereby reducing the number of required coordination updates without additional subsystem solving steps. These results show that Anderson prediction substantially reduces the coordination effort and computational time required by BGEP-D while preserving the alternating EPS--DHN coordination structure.
	
	\begin{figure}[H]
		\centering
		\includegraphics[width=0.9\columnwidth]{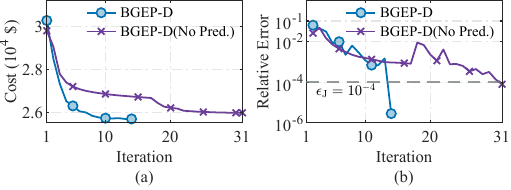}
		\vspace{-0.5em}
		\caption{Stage-II convergence trends of BGEP-D with and without Anderson prediction on the E6-H6 IHPS: (a) operating costs and (b) relative cost errors.}
		\label{fig4:caseA_pc_ablation}
		\vspace{-0.3em}
	\end{figure}
	
	\standalonesubsubsection{Performance Under Different DHN Operation Modes}
	Table~\ref{tab:variable_operation_modes} compares BGEP-D and NMSE-D under the CF-VT and VF-CT operation modes. Together with the VF-VT case presented above, these cases illustrate the performance attained by BGEP-D under different DHN operation modes. For BGEP-D, the incurred operating cost increases from \$25,684.7 under VF-VT to \$27,595.75 under CF-VT, where the fixed mass flow limits the flexibility of heat transport and CHP operation. The cost further increases to \$30,773.51 under VF-CT, indicating a stronger restriction on CHP operation with a fixed heat-source supply temperature. Similar cost variations are observed for NMSE-D, indicating that the differences across the three cases mainly result from different DHN operation modes.
	
	Under the CF-VT operation mode, the prescribed mass flow keeps the DHN equivalent fixed throughout the dispatch process, and the decentralized procedure requires only a single DHN equivalent projection without iterative EPS--DHN coordination. BGEP-D reduces the operating cost by 3.0\% and the wind curtailment from 12.84 to 3.11~MWh. Under the VF-CT operation mode, the supply temperature of the heat source is prescribed while the mass flow remains variable, and the DHN equivalent is therefore updated with the mass flow during coordination. BGEP-D completes Stage II in two iterations, compared with ten iterations for NMSE-D, while reducing the operating cost by 3.65\% and accommodating all available wind power generation, whereas NMSE-D results in 25.64~MWh of wind curtailment. Together with the VF-VT results, these comparisons show that BGEP-D remains applicable under different DHN operation modes and maintains its economic and dispatch advantages despite the DHN flexibility restrictions.
	
	\vspace{-0.3em}
	\begin{table}[H]
		\caption{Performance Under Different DHN Operation Modes}
		\label{tab:variable_operation_modes}
		\centering
		\footnotesize
		\setlength{\tabcolsep}{1.5pt}
		\renewcommand{\arraystretch}{1.04}
		\begin{tabular*}{\columnwidth}{@{\extracolsep{\fill}}cccccc@{}}
			\toprule
			Mode & Method 
			& \makecell[c]{Operating\\Cost (\$)} 
			& \makecell[c]{Wind Curtailment\\ (MWh)} 
			& \makecell[c]{Iterations\\(Stages I+II)}
			& \makecell[c]{Runtime\\(s)} \\
			\midrule
			\multirow[c]{2}{*}{CF--VT} & BGEP-D & 27,595.75 & 3.11 & -- & 0.097 \\
			& NMSE-D & 28,448.09 & 12.84 & -- & 0.010 \\
			\midrule
			\multirow[c]{2}{*}{VF--CT} & BGEP-D & 30,773.51 & 0.00 & 2$+$2 & 2.589 \\
			& NMSE-D & 31,940.23 & 25.64 & 2$+$10 & 3.252 \\
			\bottomrule
		\end{tabular*}
	\end{table}
	
	\subsection{Simulation on a Large-Scale IHPS}
	The economic and computational results obtained for the large-scale E39--H20 IHPS are summarized in Table~\ref{tab:caseB_large_scale_performance}, while Fig.~\ref{fig5:caseB_performance} compares the Stage-II operating costs and average closure gaps of the different approaches. Among the methods considered, BGEP-D achieves the lowest operating cost and shortest total runtime. Compared with NMSE-D, BGEP-D reduces the operating cost by 3.59\% and shortens the total runtime from 69.10~s to 17.32~s. With comparable numbers of Stage-II iterations, the runtime difference is mainly reflected in the DHN subproblem, which requires 16.65~s for BGEP-D and 68.91~s for NMSE-D. This indicates that the computational improvement primarily results from the reduced effort required to solve the nonlinear DHN subproblem at each iteration.
	
	Fig.~\ref{fig5:caseB_performance} further shows that BGEP-D maintains a lower operating cost throughout Stage II and a zero average closure gap at the reported precision level, whereas the NMSE-D gap remains above 1.4 $^\circ{\rm C}$ after the initial iteration. This persistent gap reflects the different thermal representations that are employed by the DHN equivalent and subproblem of NMSE-D, which reconstruct different internal thermal states. In contrast, the DHN equivalent and subproblem of BGEP-D are derived from the same Bernstein--Galerkin thermal relations, maintaining thermal consistency as the mass flow trajectory is updated. These results demonstrate that BGEP-D retains its economic and computational advantages while maintaining thermal consistency at a larger network scale.
	
	The effect of Anderson prediction is further evaluated by comparing BGEP-D with its counterpart without prediction. Both cases terminate after five Stage-II iterations, while the prediction method reduces the time required to solve the DHN subproblem from 50.87~s to 16.65~s and the total runtime from 51.73~s to 17.32~s, with a marginal difference between their operating costs. The same number of Stage-II iterations, together with the decrease in the average DHN solution time, indicates that the prediction process changes the coordination sequence and the associated nonlinear DHN subproblems, thereby reducing their required solution effort.
	
	\begin{table}[H]
		\caption{Economic and Computational Performance on the E39-H20 IHPS}
		\label{tab:caseB_large_scale_performance}
		\centering
		\footnotesize
		\setlength{\tabcolsep}{1.3pt}
		\renewcommand{\arraystretch}{1.05}
		\begin{tabular*}{\columnwidth}
			{@{\extracolsep{\fill}}lccccc@{}}
			\toprule
			\multirow{2}{*}[-0.5ex]{Method}
			& \multirow{2}{*}[-0.5ex]{\makecell[c]{Operating Cost\\(\$)}}
			& \multirow{2}{*}[-0.5ex]{\makecell[c]{Iterations\\(Stages I$+$II)}}
			& \multicolumn{3}{c}{Runtime (s)} \\
			\cmidrule(lr){4-6}
			& & & EPS & DHN & Total \\
			\midrule
			BGEP-D
			& 1,690,999.33 & 2$+$5 & 0.67 & 16.65 & 17.32 \\
			\makecell[l]{BGEP-D\\(No Pred.)}
			& 1,691,165.42 & 2$+$5 & 0.86 & 50.87 & 51.73 \\
			BGM-C
			& 1,746,164.21 & -- & -- & -- & 87.94 \\
			NMSE-D
			& 1,753,987.70 & 2$+$4 & 0.19 & 68.91 & 69.10 \\
			NM-C
			& 1,759,489.72 & -- & -- & -- & 263.58 \\
			\bottomrule
		\end{tabular*}
		\vspace{-0.4em}
	\end{table}
	
	\begin{figure}[H]
		\centering
		\includegraphics[width=0.9\columnwidth]{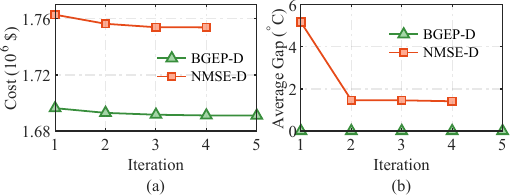}
		\vspace{-0.5em}
		\caption{Stage-II performance of BGEP-D and NMSE-D on the E39--H20 IHPS: (a) operating costs and (b) average closure gaps.}
		\label{fig5:caseB_performance}
	\end{figure}
	
	\endgroup
	
	\section{Conclusion}
	A continuous-time decentralized power dispatch framework for IHPSs implemented under the VF-VT operation mode is developed on the basis of a Bernstein--Galerkin equivalent projection and Anderson prediction. The DHN equivalent and subproblem share a unified Bernstein-space formulation that preserves continuous spatiotemporal thermal dynamics and enables EPS--DHN coordination to be performed through boundary variables without disclosing internal information. Case studies demonstrate that the proposed framework maintains thermal state consistency throughout the decentralized coordination while improving the economic and computational performance of IHPS power dispatch.
	
	Future work will focus on decentralized IHPS power dispatch under electricity and heat uncertainties, as these uncertainties can affect the modeling of thermal dynamics and the computational performance of decentralized optimization schemes.
	
	\bibliographystyle{IEEEtran}
	\bibliography{document}

@IEEEtranBSTCTL{IEEEexample:BSTcontrol,
  CTLuse_forced_etal       = "yes",
  CTLmax_names_forced_etal = "3",
  CTLnames_show_etal       = "3",
  CTLdash_repeated_names   = "no"
}

@article{khatibi2021exploiting,
  author={Khatibi, Mahmood and Bendtsen, Jan Dimon and Stoustrup, Jakob and Mølbak, Tommy},
  journal={IEEE Trans. Smart Grid}, 
  title={Exploiting Power-to-Heat Assets in District Heating Networks to Regulate Electric Power Network}, 
  year={2021},
  volume={12},
  number={3},
  pages={2048-2059},
  doi={10.1109/TSG.2020.3044348}}

@article{xue2021reconfiguration,
  author = {Xue, Yixun and Shahidehpour, Mohammad and Pan, Zhaoguang and Wang, Bin and Zhou, Quan and Guo, Qinglai and Sun, Hongbin},
  title = {Reconfiguration of District Heating Network for Operational Flexibility Enhancement in Power System Unit Commitment},
  journal = {IEEE Trans. Sustain. Energy},
  volume = {12},
  number = {2},
  pages = {1161--1173},
  year = {2021}
}

@article{cao2019decentralized,
  author  = {Cao, Yang and Wei, Wei and Wu, Lei and Mei, Shengwei and Shahidehpour, Mohammad and Li, Zhiyi},
  title   = {Decentralized Operation of Interdependent Power Distribution Network and District Heating Network: A Market-Driven Approach},
  journal = {IEEE Trans. Smart Grid},
  year    = {2019},
  volume  = {10},
  number  = {5},
  pages   = {5374--5385},
  doi     = {10.1109/TSG.2018.2880909}
}

@article{zhu2026fully,
  author = {Zhu, Haohao and Li, Jiayi and Zhu, Jizhong and Gao, Meiyun and Liao, Chenlei and Zhang, Di},
  title = {A Fully Distributed Incentive Mechanism for Integrated Electricity and Heat Systems},
  journal = {IEEE Trans. Sustain. Energy},
  volume = {17},
  number = {1},
  pages = {449--463},
  year = {2026}
}

@article{lu2020high,
  author = {Lu, Shuai and Gu, Wei and Zhou, Suyang and Yu, Wenwu and Yao, Shuai and Pan, Guangsheng},
  title = {High-Resolution Modeling and Decentralized Dispatch of Heat and Electricity Integrated Energy System},
  journal = {IEEE Trans. Sustain. Energy},
  volume = {11},
  number = {3},
  pages = {1451--1463},
  year = {2020}
}

@article{du2025globally,
  author = {Du, Yuan and Xue, Yixun and Shahidehpour, Mohammad and Wu, Wenchuan and Chang, Xinyue and Li, Zening and Sun, Hongbin},
  title = {Globally Optimal Distributed Operation of Integrated Electric and Heating Systems},
  journal = {IEEE Trans. Sustain. Energy},
  volume = {16},
  number = {1},
  pages = {336--349},
  year = {2025}
}

@article{zheng2021dynamic,
  author = {Zheng, Weiye and Hou, Yunhe and Li, Zhigang},
  title = {A Dynamic Equivalent Model for District Heating Networks: Formulation, Existence and Application in Distributed Electricity-Heat Operation},
  journal = {IEEE Trans. Smart Grid},
  volume = {12},
  number = {3},
  pages = {2685--2695},
  year = {2021}
}

@article{qiu2023decentralized,
  author = {Haifeng Qiu and Ashwin Vinod and Shuai Lu and Hoay Beng Gooi and Guangsheng Pan and Suhan Zhang and Veerapandiyan Veerasamy},
  title = {Decentralized Mixed-Integer Optimization for Robust Integrated Electricity and Heat Scheduling},
  journal = {Appl. Energy},
  volume = {350},
  year = {2023},
  note = {{Art. no.} 121693}
}

@article{chen2021integrated,
  author  = {Chen, Yuwei and Guo, Qinglai and Sun, Hongbin and others},
  title   = {Integrated Heat and Electricity Dispatch for District Heating Networks With Constant Mass Flow: A Generalized Phasor Method},
  journal = {IEEE Trans. Power Syst.},
  year    = {2021},
  volume  = {36},
  number  = {1},
  pages   = {426--437},
  doi     = {10.1109/TPWRS.2020.3008345}
}

@article{qin2022increasing,
  author = {Qin, Xin and Guo, Ye and Shen, Xinwei and Sun, Hongbin},
  title = {Increasing Flexibility of Combined Heat and Power Systems Through Optimal Dispatch With Variable Mass Flow},
  journal = {IEEE Trans. Sustain. Energy},
  volume = {13},
  number = {2},
  pages = {986--997},
  year = {2022}
}

@article{wu2022distributionally,
  author = {Xuewei Wu and Jiakun Fang and Zhe Chen},
  title = {Distributionally Robust Unit Commitment of Integrated Electricity and Heat System Under Bi-Directional Variable Mass Flow},
  journal = {Appl. Energy},
  volume = {326},
  year = {2022},
  note = {{Art. no.} 119788}
}

@article{huang2017coordinated,
  author = {Jinbo Huang and Zhigang Li and Q.H. Wu},
  title = {Coordinated Dispatch of Electric Power and District Heating Networks: A Decentralized Solution Using Optimality Condition Decomposition},
  journal = {Appl. Energy},
  volume = {206},
  pages = {1508--1522},
  year = {2017}
}

@article{xue2020coordinated,
  author = {Xue, Yixun and Li, Zhengshuo and Lin, Chenhui and Guo, Qinglai and Sun, Hongbin},
  title = {Coordinated Dispatch of Integrated Electric and District Heating Systems Using Heterogeneous Decomposition},
  journal = {IEEE Trans. Sustain. Energy},
  volume = {11},
  number = {3},
  pages = {1495--1507},
  year = {2020}
}

@article{tan2024parti,
  author = {Tan, Zhenfei and Yan, Zheng and Zhong, Haiwang and Xia, Qing},
  title = {Non-Iterative Solution for Coordinated Optimal Dispatch via Equivalent Projection--Part {I}: Theory},
  journal = {IEEE Trans. Power Syst.},
  volume = {39},
  number = {1},
  pages = {890--898},
  year = {2024}
}

@article{tan2024partii,
  author = {Tan, Zhenfei and Yan, Zheng and Zhong, Haiwang and Xia, Qing},
  title = {Non-Iterative Solution for Coordinated Optimal Dispatch via Equivalent Projection--Part {II}: Method and Applications},
  journal = {IEEE Trans. Power Syst.},
  volume = {39},
  number = {1},
  pages = {899--908},
  year = {2024}
}

@article{zheng2021noniterative,
  author = {Zheng, Weiye and Wu, Wenchuan and Li, Zhigang and Sun, Hongbin and Hou, Yunhe},
  title = {A Non-Iterative Decoupled Solution for Robust Integrated Electricity-Heat Scheduling Based on Network Reduction},
  journal = {IEEE Trans. Sustain. Energy},
  volume = {12},
  number = {2},
  pages = {1473--1488},
  year = {2021}
}

@article{zheng2022distributed,
  author = {Zheng, Weiye and Hill, David J.},
  title = {Distributed Real-Time Dispatch of Integrated Electricity and Heat Systems With Guaranteed Feasibility},
  journal = {IEEE Trans. Ind. Informat.},
  volume = {18},
  number = {2},
  pages = {1175--1185},
  year = {2022}
}

@article{zheng2023distributed,
  author = {Zheng, Weiye and Zhu, Jizhong and Luo, Qingju.},
  title = {Distributed Dispatch of Integrated Electricity-Heat Systems With Variable Mass Flow},
  journal = {IEEE Trans. Smart Grid},
  volume = {14},
  number = {3},
  pages = {1907--1919},
  year = {2023}
}

@article{yang2020equivalent,
  author = {Yang, Jingwei and Zhang, Ning and Botterud, Audun and Kang, Chongqing},
  title = {On an Equivalent Representation of the Dynamics in District Heating Networks for Combined Electricity-Heat Operation},
  journal = {IEEE Trans. Power Syst.},
  volume = {35},
  number = {1},
  pages = {560--570},
  year = {2020}
}

@article{parvania2016unit,
  author = {Parvania, Masood and Scaglione, Anna},
  title = {Unit Commitment With Continuous-Time Generation and Ramping Trajectory Models},
  journal = {IEEE Trans. Power Syst.},
  volume = {31},
  number = {4},
  pages = {3169--3178},
  year = {2016}
}

@article{zheng2021energy,
  author = {Zheng, Chao and Fang, Jiakun and Wang, Shaorong and Ai, Xiaomeng and Liu, Zhou and Chen, Zhe},
  title = {Energy Flow Optimization of Integrated Gas and Power Systems in Continuous Time and Space},
  journal = {IEEE Trans. Smart Grid},
  volume = {12},
  number = {3},
  pages = {2611--2624},
  year = {2021}
}

@article{zhou2023function,
  author = {Bo Zhou and Xiaomeng Ai and Jiakun Fang and Kun Li and Wei Yao and Zhe Chen and Jinyu Wen},
  title = {Function-Space Optimization to Coordinate Multi-Energy Storage Across the Integrated Electricity and Natural Gas System},
  journal = {Int. J. Electr. Power Energy Syst.},
  volume = {151},
  year = {2023},
  note = {{Art. no.} 109181}
}

@article{walker2011anderson,
  author = {Walker, Homer F. and Ni, Peng},
  title = {Anderson Acceleration for Fixed-Point Iterations},
  journal = {SIAM J. Numer. Anal.},
  volume = {49},
  number = {4},
  pages = {1715--1735},
  year = {2011}
}

@article{saad2025acceleration,
  author = {Saad, Yousef},
  title = {Acceleration Methods for Fixed-Point Iterations},
  journal = {Acta Numer.},
  volume = {34},
  pages = {805--890},
  year = {2025}
}

@misc{deng2026supplementary,
  author = {Deng, Jie.},
  title = {Detailed Case Data},
  year = {2026},
  note = {[Online]. Available: \url{https://github.com/jiezi95/Supplementary-Models-and-Case-Data}}
}

@ARTICLE{li2015ComebinedHeat,
  author={Li, Zhigang and Wu, Wenchuan and Shahidehpour, Mohammad and Wang, Jianhui and Zhang, Boming},
  journal={IEEE Trans. Sustain. Energy}, 
  title={Combined Heat and Power Dispatch Considering Pipeline Energy Storage of District Heating Network}, 
  year={2016},
  volume={7},
  number={1},
  pages={12-22}
  }

@ARTICLE{deng2026integrated,
  author  = {Jie Deng and Zhigang Li and J. H. Zheng and Ye Guo},
  title   = {Integrated Heat and Power System Scheduling with Continuous-Time Thermal Dynamics via {Bernstein-Galerkin} Optimization},
  journal = {IEEE Trans. Smart Grid},
  year    = {2026},
  note    = {accepted. Available: \url{https://arxiv.org/abs/2608.17287}}
}
	
\end{document}